\documentclass[twocolumn]{aastex63}
\usepackage{amsmath}
\usepackage{amssymb}
\usepackage{multirow,booktabs}

\renewcommand{\figurename}{Figure }
\renewcommand{\tablename}{Table }
\newcommand{\eqnname}{Equation }
\newcommand{\secname}{Section }
\newcommand{\midplane}{mid-plane }
\newcommand{\Gaia}{\emph{Gaia }}

\received{XXX}
\revised{XXX}
\accepted{XXX}
\renewcommand{\today}{March 11, 2022}

\shorttitle{Sgr--Milky-Way-Disc Interaction}
\shortauthors{Bennett, Bovy, \& Hunt}
\graphicspath{{./}{plots/}}

\begin{document}

\title{Exploring the Sgr--Milky-Way-disc interaction using high resolution $N$-body simulations}

\correspondingauthor{Jo Bovy}
\email{bovy@astro.utoronto.ca}

\author{Morgan Bennett}
\affiliation{Department of Astronomy and Astrophysics, University of Toronto, 50 St. George Street, Toronto ON, M5S 3H4, Canada}

\author[0000-0001-6855-442X]{Jo Bovy}
\affiliation{Department of Astronomy and Astrophysics, University of Toronto, 50 St. George Street, Toronto ON, M5S 3H4, Canada}

\author[0000-0001-8917-1532]{Jason A. S. Hunt}
\affiliation{Center for Computational Astrophysics, Flatiron Institute, 162 5th Av., New York City, NY 10010, USA}

\begin{abstract}
The ongoing merger of the Sagittarius (Sgr) dwarf galaxy with the Milky Way is believed to strongly affect the dynamics of the Milky Way's disc. We present a suite of 13 $N$-body simulations, with 500 million to 1 billion particles, modelling the interaction between the Sagittarius dwarf galaxy (Sgr) and the Galactic disc. To quantify the perturbation to the disc's structure and dynamics in the simulation, we compute the number count asymmetry and the mean vertical velocity in a solar-neighbourhood-like volume. We find that, overall, the trends in the simulations match those seen in a simple one-dimensional model of the interaction. We explore the effects of changing the mass model of Sgr, the orbital kinematics of Sgr, and the mass of the Milky Way halo. We find that none of the simulations match the observations of the vertical perturbation using \Gaia Data Release 2. In the simulation that is the most similar, we find that the final mass of Sgr far exceeds the observed mass of the Sgr remnant, the asymmetry wavelength is too large, and the shape of the asymmetry does not match past $z\approx0.7$ kpc. We therefore conclude that our simulations support the conclusion that Sgr alone could not have caused the observed perturbation to the solar neighbourhood.
\end{abstract}

\keywords{Galaxy: disc --- Galaxy: kinematics and dynamics --- Galaxy: structure --- Galaxy: evolution --- solar neighbourhood --- Galaxy: formation}

\section{Introduction}\label{sec:intro}

The Milky Way has been the ideal location to study galactic dynamics because our position inside the Galactic Disc gives us the perfect view of stars' motions. With surveys like \Gaia Data Release 2 (DR2; \citealt{Gaia}) we have an unprecedented amount of data on the dynamics of stars in the solar neighbourhood. In particular, there has been a burst of interest in the out-of-equilibrium dynamics of the disc in the solar neighbourhood \citep[e.g.,][]{antoja,Kawata2018,Bennett2019,BH-GALAH}.

Oscillations in the solar neighbourhood were first discovered by \citet{widrow12} by looking at the number count asymmetry and trends in the mean vertical velocity. They used a simple one-dimensional $N$-body simulation to show that modes of the type observed were in fact a natural occurrence in the vertical disc structures. After the discovery of the solar neighbourhood oscillations, the work of \citet{Gomez2013} was one of the first to simulate the Sgr-Milky Way interaction as a possible explanation for the observations. Those authors focused on the impacts of Sgr across the disc using a light ($10^{10.5}$ M$_\odot$) and a heavy ($10^{11}$ M$_\odot$) Sgr model. They were able to achieve a good match to the mean vertical velocity measurements, but found that their asymmetry wavelength was larger than perturbations by a factor than two, a common discrepancy with Sgr simulations \citep{Laporte2018}.

Using simulations to investigate the cause of the local perturbation became an important tool for the field. \citet{Donghia2016} looked at not only one large satellite as the possible perturber, but also a collection of large satellites moving at different velocities as predicted by cosmological simulations. They found that a singular massive satellite would cause underdense regions to experience coupled horizontal and vertical oscillations of stars. The simulation with multiple large satellites led to a wobble in the disc as well as flares in the outer disc and gradual heating. The idea of looking at the effects of more than just one perturber was furthered by \citet{Chequers2018} who looked at the effects of halo substructure such as dark matter subhaloes on the vertical profile of the disc. Their simulations showed that the existence of subhaloes can excite pre-existing modes present in the disc. They theorized that the oscillations in the disc may be the result of a dynamically active disc and not necessarily the passage of a large perturber.

One of the most widely used Sgr-Milky Way simulations to date, \citet{Laporte2018} explores how the force from the Sagittarius Dwarf Galaxy (Sgr) and the Large Magellanic Clouds (LMC) affects the equilibrium of the Milky Way. They look beyond just the one-dimensional vertical perturbation and also consider the effect as a function of radius, finding that Sgr causes corrugation and flaring of the disc. 

With \Gaia DR2, observations of the perturbation improved by leaps and bounds. \citet{Bennett2019} were able to utilize the completeness of \Gaia DR2 to improve measurements of the number count asymmetry and the large radial velocity sample in \Gaia DR2 to improve measurements of the mean vertical velocity. Furthermore, \citet{antoja} discovered the existence of a phase-space spiral, the two-dimensional projection of the same effect. \citet{antoja} also used a small test particle simulation to show that the phase-space spiral could be reproduced by phase mixing after a perturbation to the disc. The effects due to Sgr in the \citet{Laporte2018} simulation were then further analysed in \citet{Laporte2019} where they compared it to the observed phase-space spiral. They found that there is a qualitative match, but were not able to reproduce the perturbation wavelength with their simulations. Further, while they consider four mass models for Sgr in the first paper, the second paper only focused on one orbit for one of the lighter Sgr models. 

Once again considering alternatives to Sgr as the cause of the perturbation, \citet{Khoperskov2019} used an $N$-body simulation to show that the bar buckling could reproduce a qualitative match to the vertical phase-space perturbation. However, while they found they were able to relax the timing of the perturbation, the timing of the perturbation may still require recurring buckling instead of a one-time event to explain the perturbation.

Most recently, \citet{BH2021} performed a high-resolution ($10^8$ particles) $N$-body simulation that modelled the interaction between an impulsive mass (Sgr was represented by a point mass) and a cold stellar disc. They found that the resultant perturbation from the interaction resulted in a density wave and a bending wave that can survive for 1.5 Gyr. They also found that to match both the current mass of Sgr as well as the amplitude of the perturbation, they required that the observed oscillations must have been excited by Sgr on a previous pericentric passage $\sim1-2$ Gyr ago. They also required that Sgr be losing mass at a rate of 0.5-1 dex per orbit, which is fairly high \citep{Tollet2017}. Finally, in two of their twelve considered volumes, they were better able to reproduce the tightness of the spiral in the radial velocity but again found that the azimuthal velocity phase spiral was too loosely wound when compared to observations. However, their use of a point mass for Sgr meant that comparing the final position of the point mass to the true position of Sgr was not feasible and also ignored mass loss of Sgr as it passed through a pericentre.

The focus of simulations thus far has been on making qualitative comparisons to the solar neighbourhood observations. We aim to produce simulations that can be used to quantitatively compare the effects of Sgr to the \Gaia DR2 observations. While there exists a large number of simulations on the interaction between a massive perturber and the Galactic disc, the varying simulation methods discussed above mean it is difficult to directly compare the properties of the solar neighbourhood in each. By utilising a fast GPU $N$-body tree-code, we are able to overcome this and quickly run multiple simulations with 500 million to 1 billion particles \citep[e.g.][]{Hunt2021}. We choose the simulations to run based on our previous investigation of Sgr--Milky-Way interactions and whether they can reproduce the observed phase-space structure in the solar neighbourhood \citep{Bennett2021}. In that paper, we used a linear-perturbation-theory model to determine the impact of Sgr on the kinematics of the solar neighbourhood. \citet{Vasiliev2020} performed an in-depth analysis of the kinematics of Sgr including estimating the mass of the remnant, $\sim (4\pm1)\times10^8\,\mathrm{M_\odot}$, where approximately 25\% comes from the stellar component. They also investigated how the initial structure of Sgr relates to the observed remnant today and were able to reproduce the cigar-like shape using spherical models. The Sgr mass models used in our analysis are all based on the models considered in \citet{Vasiliev2020}.

In \secname \ref{sec:numericalmethods}, we discuss the techniques we use to initialise the initial conditions for our three-dimensional $N$-body simulations. We also discuss how we choose where to place Sgr in our Milky Way for the simulations and we include a brief discussion about the properties of our equilibrium simulations before placing Sgr. In \secname \ref{sec:solarneighbourhood}, we discuss the properties of solar neighbourhood-like volumes in each of our simulations. That \secname is broken down into three further subsections where we look at the effects of changing (i) the mass of Sgr, (ii) Sgr's orbit, and (iii) the mass of the Milky Way halo. Finally, in \secname \ref{sec:conc} we discuss our findings, their implications, and how to move forward. 

\section{Numerical Methods}\label{sec:numericalmethods}

\begin{deluxetable}{ccccc}
\tablecaption{Milky Way Halo parameters for the three equilibrium initial conditions.}\label{tbl:MWhalo}
\tablecolumns{5}
\tablewidth{0pt}
\tablehead{\colhead{} &
  \colhead{$M_{200}$} &
  \colhead{$c_{200}$} & 
  \colhead{$M_h$} & 
  \colhead{$a_h$}\\
  \colhead{} & 
  \colhead{$\left(10^{12}\, \mathrm{M}_\odot\right)$} &
  \colhead{} &
  \colhead{$\left(10^{12}\, \mathrm{M}_\odot\right)$} & 
  \colhead{$\left(\mathrm{kpc}\right)$}}
\startdata
 Light   & 0.6    & 8.55 & 0.8 & 26.4 \\
 Medium  & 1.0   & 10.10 & 1.2 & 27.7 \\
 Heavy   & 1.4  & 11.35 & 1.6 & 28.6
\enddata
\end{deluxetable}

To create the initial conditions for our $N$-body simulation, we use \texttt{GALIC: Galaxy initial conditions construction}\footnote{\url{https://wwwmpa.mpa-garching.mpg.de/~volker/galic/}}\citep{Yurin2014} to set up the Milky Way and Sgr, separately. \texttt{GalIC} uses the minimization of a merit function to solve the collisionless Boltzmann equation by adjusting the particle velocities. It randomly draws new velocities from an approximate distribution function and then only keeps those that improve the fit of the merit function. We then simulate the evolution of the Milky-Way--Sgr system using \texttt{Bonsai}\footnote{\url{https://github.com/treecode/Bonsai}} \citep{Bedorf12}. In this section, we discuss all of the ingredients of our numerical methods in detail. 

\subsection{Model for the Milky Way}\label{sec:MWmodel}

For the purposes of our investigation, we choose to look at three different Milky Way models. The models are derived from the \texttt{galpy} potential \texttt{MWPotential2014} \citep{galpy}, but with heavier halos and they match the potentials in \citet[][hereafter BB21]{Bennett2021}: MWP14-1, MWP14-2, and MWP14-3. In BB21, we showed that the heavier Milky Way potentials were the only two that led to realistic looking asymmetries. However, for the sake of completeness, we also run simulations for the MWP14-1 model. All three potential models have the same disc and bulge as \texttt{MWPotential2014}, but while MWP14-1 also has the same halo as \texttt{MWPotential2014}, MWP14-2 and MWP14-3's halos are 1.5 and 2 times heaver than the \texttt{MWPotential2014} one, respectively. Their mass and concentration are shown in \tablename \ref{tbl:MWhalo} using a Hubble constant of $100\,\, \mathrm{km\,s^{-1}\,Mpc^{-1}}$ to remain consistent with the internal units of \texttt{GalIC}. Our initial setup of MWP14-2 contains approximately one billion particles while our initial conditions for MWP14-1 and MWP14-3 contains approximately 500 million particles. The number of particles in each component is approximately divided into 50\% in the disc, 10\% in the bulge, and the remaining 40\% in the halo. For the purpose of this section, the conversion between the Hernquist parameters and measured observational parameters of the Milky Way and the corresponding \texttt{GalIC} parameters will be demonstrated for MWP14-2. The method was the same for all three equilibrium initial conditions and the final \texttt{GalIC} parameters for all three Milky Ways are shown in \tablename \ref{tbl:MWGalIC}.

\tabletypesize{\scriptsize}
\begin{deluxetable}{cccc}
\tablecaption{GalIC parameters to generate the initial conditions for the three different Milky Way models.}\label{tbl:MWGalIC}
\tablecolumns{4}
\tablewidth{0pt}
\tablehead{\colhead{Parameters} &
  \colhead{MWP14-1} &
  \colhead{MWP14-2} & 
  \colhead{MWP14-3}}
\startdata
        Halo concentration $c$ (CC) & 9.0139 & 10.4295 & 11.6131 \\
        Virial velocity $v_{200}$ (V200; $\mathrm{km\,s}^{-1}$) & 142.0 & 165.4 & 184.7 \\
        Spin parameter $\lambda$ (LAMBDA) & 0.0526 & 0.0468 & 0.0441 \\
        Disc mass fraction $m_d$ (MD) & 0.1002 & 0.0636 & 0.0458\\
        Bulge mass fraction $m_b$ (MB) & 0.00676 & 0.00428 & 0.00307\\
        Disc spin fraction $j_d$ (JD) & 0.1002 & 0.0636 & 0.0458 \\
        Disk scale height (DiskHeight) & 0.0933 & 0.0933 & 0.0933\\
        Bulge Size (BulgeSize) & 0.0203 & 0.01936 & 0.0188\\
        Halo shape parameter (HaloStretch) & 1.0 & 1.0 & 1.0\\
        Bulge shape parameter (BulgeStretch) & 1.0 & 1.0 & 1.0\\ \hline
        $\mathrm{N_{HALO} }$ & 40,000,000 & 40,000,000 & 40,000,000\\
        $\mathrm{N_{DISK} }$ & 50,000,000 & 50,000,000 & 50,000,000\\
        $\mathrm{N_{BULGE}}$  & 10,000,000 & 10,000,000 & 10,000,000\\ \hline
        Halo velocity structure & 0 & 0 & 0\\
        Disk velocity structure & 3 & 3 & 3\\
        Bulge velocity structure & 0 & 0 & 0\\
        Disk dispersion R over Z ratio $\langle v_R^2\rangle/\langle v_z^2\rangle$ & 1.874 & 1.874 & 1.874\\
        Disk streaming velocity parameter $k$ & 0.985 & 0.985 & 0.985
\enddata
\end{deluxetable}

When calculating the \texttt{GalIC} parameters, the first consideration is that \texttt{MWPotential2014} uses an NFW halo, but \texttt{GalIC} assumes a Hernquist profile for the halo. \texttt{GalIC} uses the input velocity at the virial radius, $v_{200}$, to calculate the virial radius, $r_{200}$, for the initial conditions. To relate the two, they assume that the density profile of the halo may be approximated as an isothermal sphere. This leads to the relation \citep{Mo1998}:
\begin{equation}
r_{200}=\frac{v_{200}}{10H},
\end{equation} 
where $H$ is Hubble constant and \texttt{GalIC} uses a value of $H=100\,\,\mathrm{km\,s^{-1}\,Mpc^{-1}}$. Furthermore, $v_{200}$ is used to calculate not only the total mass of the galaxy, but also the scale radius of the halo in conjunction with the concentration. This means that $v_{200}$ is both a property of the entire galaxy as well as a property of the halo. To accommodate the contradiction in this double assignment, we need to artificially increase the concentration parameter that is only used to calculate properties of the halo, such that $a_{h}$ matches the scale radius of the halo in \texttt{MWPotential2014}. Now that we understand how \texttt{GalIC} uses the different parameters, we can start converting between NFW and Hernquist properties. We start by using the true $r_{200}$ and concentration, $c$, of the NFW halo in \texttt{MWPotential2014} to calculate the desired Hernquist scale radius using
\begin{equation}
    a_h= \frac{r_{200}}{c}\sqrt{2\left[\ln(1+c)-\frac{c}{1+c}\right]}
    \label{eq:ah}
\end{equation}
For MWP14-2, this gives us a Hernquist scale radius of $a_h=27.68$ kpc. However, the $r_{200}$ of the halo is \emph{not} the same as the $r_{200}$ parameter in \texttt{GalIC}. Therefore, we use the same equation above along with the newly calculated scale radius of $a_h=27.68$ kpc and $r_{200}$ which corresponds to the mass of the entire galaxy to calculate the true concentration. This artificially increases the concentration parameter passed to \texttt{GalIC}.

We find the disc and bulge mass fraction simply by calculating $M_d(R=r_{200},z=r_{200})$ and $M_b(r=r_{200})$ and we divide by the total mass of all three components at $r=r_{200}$. Calculating these values for MWP14-2 gives $m_d=0.0636$ and $m_b=0.00428$. 

The next parameter we calculate is the disc spin parameter, $\lambda$:
\begin{align}
\lambda= \frac{J_d}{j_d}\frac{(f_c/2)^{1/2}}{G^{1/2}\,{M_{200}}^{3/2}\,{r_{200}}^{1/2}}
\end{align}
where $J_d$ is the disc angular momentum, $j_d = J_d/J$ is the disc's angular momentum relative to the total angular momentum (the disc spin fraction), and 
\begin{equation}
    f_c=\frac{c\left(1-\frac{1}{(1+c)^2}-\frac{2\ln(1+c)}{(1+c)}\right)}{2\left(\ln(1+c)-\frac{c}{(1+c)}\right)^2}
\end{equation} \citep{Springel1999}. Assuming conservation of momentum of the material that forms the disc, we set $j_d$ equal to the disc mass fraction $m_d$. We then find that $\lambda =0.0468$ for MWP14-2. 

In \texttt{GalIC} you can set a starting point for the disc scale height and the initial guess for the scale radius of the disc is determined by $\lambda,\,c$ and $r_{200}$. The final value for both the scale height and radius is ultimately determined iteratively using the disc momentum. The disc scale height is given in units of disc scale length, so we use the ratio of the disc scale height, $b$, and scale length, $a$, of the Miyamoto-Nagai disc potential \citep{MiyamotoNagai1975} in \texttt{MWPotential2014} to estimate this value to be 0.0933. Ultimately, \texttt{GalIC} settles on values of $H=3.41$ kpc for the scale radius and $z_h=0.318$ kpc for the scale height of the disc. 

The scale radius of the bulge is given in units of the halo scale length; the properties of MWP14-2 result in a bulge scale length of approximately $0.01936$. For both the bulge and the halo, we choose to have a spherically symmetric and isotropic velocity structure. The shape parameters of the bulge and halo are therefore one and we choose \texttt{GalIC}'s velocity structure zero for both of them.

For the disc velocity structure, we choose an axisymmetric disc, described by a distribution function of the form $f(E, Lz, I_3)$ which requires $\langle v_z^2\rangle/\langle v_R^2\rangle$, and the net rotation specified (\texttt{GalIC} velocity structure three). Since these parameters are not considered in MWP14-2 which is a static potential, we use the \emph{Gaia} DR2 data in the solar neighbourhood to estimate both. For the velocity dispersion ratio, we find a value of approximately $\langle v_R^2\rangle/\langle v_z^2\rangle=1.874$. This is approximately consistent with \citet{Mackereth2019} who found a value of $\sigma_R/\sigma_z=1.56\pm0.10$. The net rotation is given by the disc streaming parameter, $k$, which is calculated using
\begin{equation}
    k^2=\frac{\langle v_\phi\rangle^2}{\langle{v_\phi}^2\rangle-\sigma_R^2}
\end{equation}
Again using \emph{Gaia} DR2, we find $k=0.985$ in the solar neighbourhood. 

Due to computational constraints, instead of running one initial condition with $10^9$ particles, we ran 10 (or 5 in the case of MWP14-1 and MWP14-3) initial conditions with $10^8$ particles using random seeds 1000 through 10,000. While each of the ten snapshots is in equilibrium, they are not necessarily in equilibrium with each other. This is not a concern, because as we describe in \secname \ref{sec:sims}, we run all conditions together for 3 Gyr before adding in Sgr. After combining the different conditions, we compared the density profile of all three components to the density profiles from \texttt{MWPotential2014} and found that they were consistent at all distances out to approximately 500 kpc. 

\subsection{Models for Sgr}\label{sec:Sgrmodel}

The next step is choosing the models for the Sgr-like satellite in our simulation. We choose to look at the Sgr 1, Sgr 2, and Sgr 3 models from BB21 which are the three heaviest of the five considered. These models are derived from the Sgr parameters in \citet{Vasiliev2020}. Though they were ruled out in BB21, we choose the heaviest models because we wanted to generate the largest signal possible in our $N$-body simulation to overcome the uncertainties that come with having a small number of particles in the solar neighbourhood. The three Sgr mass models have halo masses of $\mathrm{M_{sgr,h}}=(50,10,5)\times 10^{9}\,\mathrm{M_\odot}$ and stellar masses of $\mathrm{M_{sgr,*}}=(1,0.2,0.1)\times 10^{9}\,\mathrm{M_\odot}$. Both components are given as Hernquist potentials with scale radii of $\mathrm{a_{sgr,h}}=(6.7, 3.0, 2.1)\,\mathrm{kpc}$ and $\mathrm{a_{sgr,*}}=(1.45,0.65,0.46)\,\mathrm{kpc}$. The total mass of each Sgr mass model is given by $M_\mathrm{init}=\mathrm{M_{sgr,h}}+\mathrm{M_{sgr,*}}$. These two-component spherical models are consistent with the ones used by \citet{Vasiliev2020} that resulted in a match to the observed Sgr kinematics.

Like with the Milky Way, we also use \texttt{GalIC} to create the initial conditions for Sgr. By setting $M_{200}=M_\mathrm{init}$, GalIC also sets $V_{200}$ and $R_{200}$ for the entire system. To make sure that the halo scale radius in GalIC matches the values above, we have to choose a concentration using \eqnname (\ref{eq:ah}), therefore we set $c_\mathrm{sgr,h}=17.96,25.24,29.15$. Our final choice when initializing Sgr with GalIC is the number of particles in each of the two components. This choice is fairly arbitrary, so we choose a number of particles in the Sgr dark matter halo such that their mass approximately matches the mass of the halo particles in our Milky Way, $m_\mathrm{mw,h}\approx 2451\,\mathrm{M_\odot}$. For the stellar particles in Sgr, we choose a number of particles such that they will have a similar mass to the particles in the Milky Way disc, $m_\mathrm{mw,d}= 133.8\,\mathrm{M_\odot}$. So this means that the halo has $N_\mathrm{sgr,h}\approx (20.4,4.08,2.04)\times10^6$ and the stellar component has $N_\mathrm{sgr,*}\approx (7.5, 1.5,  0.75)\times10^6$. Once we have generated the particles for Sgr, we have to decide where to place them relative to the centre of the Milky Way such that they end up near Sgr's location today. 

\subsection{Simulations}\label{sec:sims}

The simulations are evolved using a gravitational $N$-body GPU tree code integrator \texttt{Bonsai} \citep{Bedorf12}. There are five parameters that need to be specified for a simulation using \texttt{Bonsai}. For the opening angle we used a value of 0.4, which is typical, if not smaller than other simulations in the field \citep{Laporte2018,Khoperskov2019}. With \texttt{Bonsai}, you can specify the frequency with which you want to rebuild the tree that computes the forces, we update the tree at each step. \texttt{Bonsai} requires a constant integration time step, for all simulations discussed in this paper we use 9.778145 kyr. For the simulations where we vary the mass of Sgr, the output time step was every 4000 steps (approximately every 39 Myr), but for the kinematics of Sgr simulations as well as varying the MW halo mass, we chose to output snapshots every 1000 steps (approximately every 9.8 Myr). Finally, we use a softening parameter of 50 pc, which is on the order typically used for these types of simulations \citep{Laporte2018,Vasiliev2020}. The entire suite of simulations took approximately 5 GPU years to run on Nvidia Telsa 32 GB V100 GPUs.

We ran the initial conditions of the Milky Way for 3 Gyr or more before adding Sgr to ensure the disc was in equilibrium as well as to get a better measure of the intrinsic uncertainty in the perturbation to the distribution function. Ideally, the perturbation function would be completely symmetric for our `equilibrium' case, but due to the discrete nature of anybody simulations and intrinsic Poisson error, that is not the case. So when we measure the asymmetry in the perturbed simulation we have a baseline to compare against.

\subsection{Validation}

Before looking at the perturbed Milky Way, we first look at the equilibrium case to see how well we can trust any asymmetry seen in our perturbed simulation. All of our initial conditions are integrated on their own for 3-5 Gyr to ensure that they have reached an equilibrium. This allows us to focus on the effects of Sgr as the dominant cause of the perturbation as opposed to being muddied by other dynamical events such as the bar formation.

The first thing we check is that the properties of our disc are reasonable compared to that of the Milky Way. We examined eight solar neighbourhood-like volumes around the equilibrium Milky Way simulation spaced $\pi/4$ radians apart and starting at $\phi=0$. They were all at a radius of 8.1 kpc from the centre of mass of the disc and bulge. The true mid-plane density has been measured to be approximately $\rho_0= 0.1$ \citep{Holmberg2000,Widmark2019}. The average \midplane density of MWP14-1 is $(0.062 \pm 0.002)$ $\mathrm{M_\odot\,pc^{-3}}$ which is significantly lower than the \midplane density in the Milky Way, which we will have to account for in our analysis. For MWP14-2, we found an average \midplane density of  $(0.116 \pm 0.006)\, \mathrm{M_\odot\,pc^{-3}}$. For MWP14-3, the average \midplane density fell at $(0.12 \pm 0.01) \, \mathrm{M_\odot\,pc^{-3}}$. The average \midplane density is high compared to true \midplane density of the Milky Way. However, the density is also affected by the introduction of Sgr, and therefore an exact match at this stage is not required. Next, we looked at the velocity dispersion of the simulated Milky Way discs and found that their mean velocity dispersion at the solar neighbourhood was $22.4 \pm 0.12\, \mathrm{km\,s^{-1}}$, $22.8 \pm 0.3\, \mathrm{km\,s^{-1}}$, and $22.3 \pm 0.4\, \mathrm{km\,s^{-1}}$ for MWP14-1, MWP14-2, and MWP14-3 respectively. This is also similar to the true values of the Milky Way, which is approximately 20.5 km\,s$^{-1}$ \citep{GaiaKinematics,Bennett2021}.

Second, we check that the Milky Way reaches an equilibrium before we add the satellite. To do this, we check the rotation curve of the galaxy at several different snapshots in the simulation. At first, the potential is smooth and well-defined. We looked at the Toomre parameter for MWP14-1, MWP14-2, and MWP14-3 of approximately $Q=1.3, 1.5$, and 1.8 respectively \citep{Toomre1964}. From these values, we would expect that they are increasingly more stable, but could still be susceptible to bar and spiral formation. As expected, after some time, all three simulations form a bar and we need to wait until the disc has reached equilibrium again before adding our satellite. 

\begin{figure}
    \centering
    \includegraphics[width=0.45\textwidth]{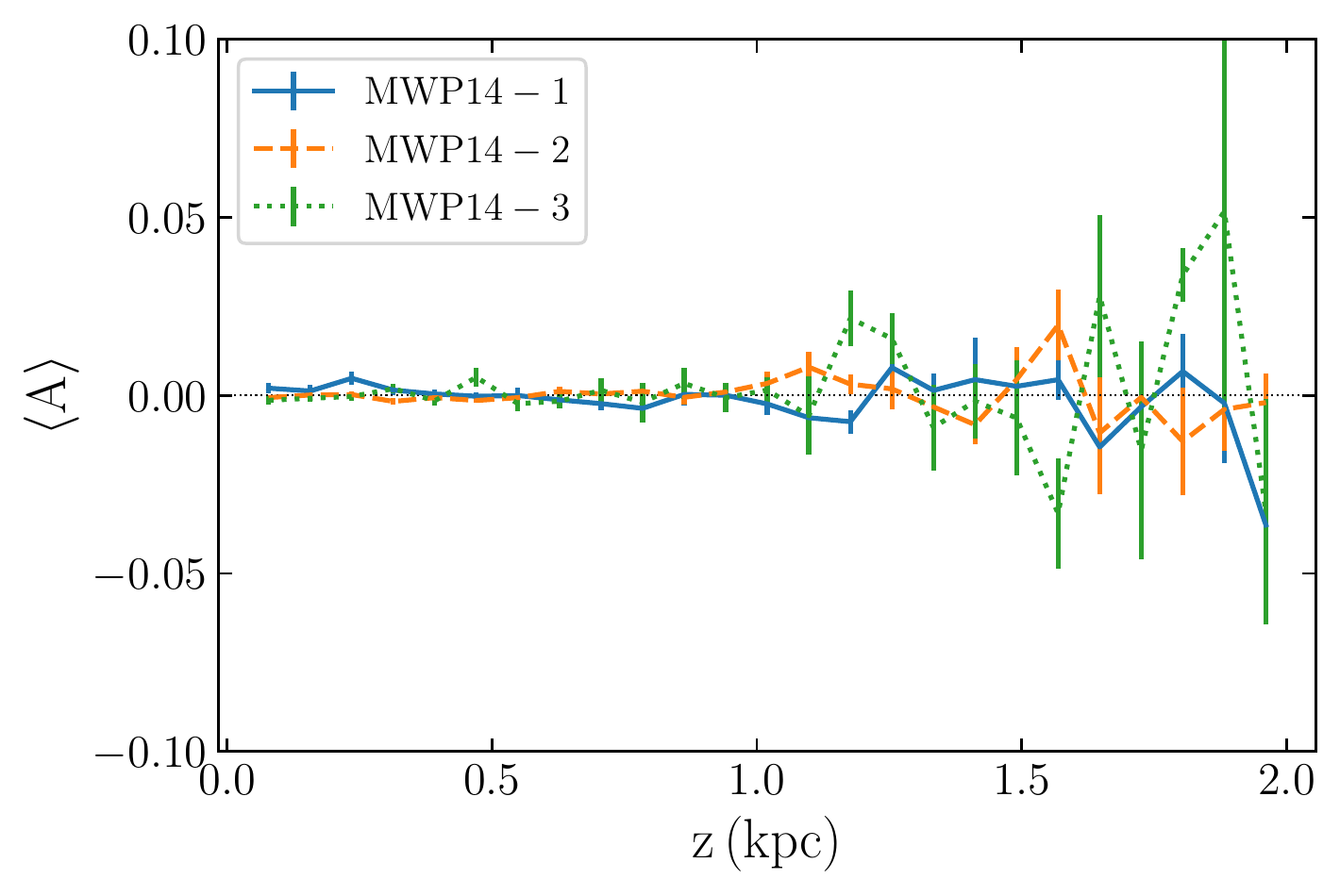}
    \caption{Mean asymmetry $A$ at eight different solar neighbourhood like areas in the simulation which are at a radius of 8.1 kpc and spaced every 45 degrees around the circle starting at $\phi=0$. The three lines represent the three different equilibrium simulations done for MWP14-1 (blue solid), MWP14-2 (orange dashed), and MWP14-3 (green dotted). This asymmetry for the unperturbed disc provides a baseline for comparison when considering the significance of the signal in our perturbed simulations.}
    \label{fig:Equil_A}
\end{figure}

Finally, we check the density asymmetry defined in Equation \eqref{eq-asymmetry} below as a function of height at our last time at several different Solar Neighbourhood-like locations throughout the Milky Way. We measure the asymmetry at the previously defined eight different equally spaced locations in the Milky Way. \figurename \ref{fig:Equil_A} shows the mean asymmetry as well as the error in the mean using those eight locations spaced equidistant around the circle at $R=8.1$ kpc. For all three Milky Ways, the uncertainty as well as the asymmetry appear to grow at approximately equal rates. The figure clearly shows that within 1.3 kpc of the mid-plane, the asymmetry for MWP14-2 and MWP14-3 is within uncertainty of zero for both equilibrium Milky Ways. However, further out we see that the asymmetry grows faster than the uncertainty. The asymmetry of MWP14-1 actually stays within uncertainty of zero except for one small deviation at $\sim0.85$ kpc at all heights. When looking at our perturbed simulations, it is important to keep in mind the heights at which we can trust our results. 

\subsection{Placing Sgr}\label{sec:placeSgr}

\tabletypesize{\small}
\begin{deluxetable*}{ccccccccccccc}
\tablecaption{Position of Sagittarius at $t=\mathrm{now}$ Compared to Each of the Simulations.}
\tablecolumns{13}
\tablewidth{0pt}
\tablehead{\colhead{} &
  \colhead{} &
  \colhead{$t$} &
  \colhead{$M_\mathrm{init}$} & 
  \colhead{$M_\mathrm{now}$} &
  \colhead{$R$} &
  \colhead{$v_R$} &
  \colhead{$v_T$} &
  \colhead{$z$} &
  \colhead{$v_z$} &
  \colhead{$\phi$} &
  \colhead{$t_\mathrm{peri}$} &
  \colhead{$r_\mathrm{peri}$} \\
  \colhead{} &
  \colhead{} &
  \colhead{(Myr)} &
  \colhead{$(10^{8}\,M_\odot)$} &
  \colhead{($10^8\,M_\odot$)} &
  \colhead{(kpc)} &
  \colhead{(km\,s$^{-1}$)} &
  \colhead{(km\,s$^{-1}$)} &
  \colhead{(kpc)} &
  \colhead{(km\,s$^{-1}$)} &
  \colhead{(rad)} &
  \colhead{(Myr)} &
  \colhead{(kpc)}}
\startdata
          \multirow{5}{*}{\rotatebox[origin=l]{90}{\begin{tabular}[c]{@{}c@{}} \bf{Sgr Mass}\\\bf{Models}\end{tabular}}} 
         & \textbf{Sgr} & \textbf{$-$} & \textbf{$-$} & $\sim 4\pm1$ & \textbf{16.7} & \textbf{227.2} & \textbf{65.8} & \textbf{-6.26} & \textbf{202.4} & \textbf{3.00} &\textbf{$-$}& \textbf{$-$}\\
        & Heavy  & 812 & 510 & 32 & 10.9 & 258.2 & 68.2 & -4.28 & 187.9 & 2.95 & 62 & 13.2\\
        & Medium  & 870 & 102 & 15 & 10.2 & 273.4 & 96.9 & -8.99 & 162.0 & 2.87 & 55 & 12.3 \\
        & Light & 909 & 51 & 10 & 13.6 & 248.8 & 76.6 & -7.59 & 186.3 & 2.95 & 55 & 8.4\\ \midrule
        \parbox[t]{2mm}{\multirow{9}{*}{\rotatebox[origin=c]{90}{\textbf{Velocity Models}}}} & \textbf{Sgr} & \textbf{$-$}& \textbf{$-$} & $\sim 4\pm1$ & \textbf{13.8} & \textbf{217.5} &  \textbf{34.1 }& \textbf{-5.52 }& \textbf{176.7 }& \textbf{2.99 } & \textbf{$-$} & \textbf{$-$}\\
        & Fastest  &  567 & 102 & 10 & 10.6 & 220.0 & 28.3 & -5.64 & 136.8 & 3.00 & 34 & 8.8 \\
        \cmidrule{2-13}
        & \textbf{Sgr} & \textbf{$-$}& \textbf{$-$} & $\sim 4\pm1$ & \textbf{15.5} & \textbf{226.3} &  \textbf{52.9 }& \textbf{-5.96 }& \textbf{191.2 }& \textbf{2.99 } & \textbf{$-$} & \textbf{$-$} \\
        & Fast  & 743 & 102 & 13 & 13.2 & 217.3 & 37.6 & -5.17 & 165.7 & 3.01 & 38 & 10.5\\ \cmidrule{2-13}
        & \textbf{Sgr} & \textbf{$-$}& $\mathbf{-}$ & $\sim 4\pm1$ & \textbf{17.8} & \textbf{225.8} & \textbf{78.8 }& \textbf{-6.56 }& \textbf{214.4 }& \textbf{3.00 } & \textbf{$-$} & \textbf{$-$}\\ 
        & Slow & 1066 & 102 & 18 & 15.5 & 224.2 & 63.5 & -5.83 & 194.5 & 3.00 & 36& 13.3 \\
        \cmidrule{2-13}
        & \textbf{Sgr} & \textbf{$-$}& $\mathbf{-}$ & $\sim 4\pm1$ & \textbf{19.8} & \textbf{231.2} & \textbf{100.6} &\textbf{ -7.06} &\textbf{ 232.2} &\textbf{ 3.01} & \textbf{$-$} & \textbf{$-$} \\
        & Slowest & 1584 & 102 & 22 & 17.0 & 233.8 & 89.5 & -6.69 & 215.8 & 2.98 & 32 & 15.6\\ \midrule
        \multirow{15}{*}{\rotatebox[origin=l]{90}{\begin{tabular}[c]{@{}c@{}} \bf{MW Mass Models}\\\bf{MWP14-3\hspace{1.4cm}MWP14-1}\end{tabular}}} & \textbf{Sgr} & \textbf{$-$}& \textbf{$-$} & $\sim 4\pm1$ & \textbf{13.8} & \textbf{217.5} & \textbf{34.1}& \textbf{-5.52}& \textbf{176.7} & \textbf{2.99 } & \textbf{$-$} & \textbf{$-$} \\
        & Fastest & 1105 & 102 & 16 & 13.5 & 217.9 & 33.5 & -4.47 & 176.5 & 3.00 & 49 & 10.3\\
        \cmidrule{2-13}
        & \textbf{Sgr} & \textbf{$-$}& \textbf{$-$} & $\sim 4\pm1$ & \textbf{16.8} & \textbf{222.2} &  \textbf{66.9} & \textbf{-6.29}& \textbf{205.0}& \textbf{3.00} & \textbf{$-$} & \textbf{$-$} \\
        & Median  & 2289 & 102 & 23 & 14.9 & 234.1 & 70.6 &  -5.92 & 198.83 & 2.98 & 42 & 13.3
 \\ \cmidrule{2-13}
        & \textbf{Sgr} & \textbf{$-$}& \textbf{$-$} & $\sim 4\pm1$ & \textbf{19.8} & \textbf{231.2} & \textbf{100.6}& \textbf{-7.06}& \textbf{232.2} & \textbf{3.01} & \textbf{$-$} &  \textbf{$-$} \\
        & Slowest  & 3707 & 102 & 40 & 19.7 & 225.6 & 97.9 & -6.14 & 232.5 & 3.01 & 43 &  17.8\\
        \cmidrule{2-13}
        & \textbf{Sgr} & \textbf{$-$}& \textbf{$-$} & $\sim 4\pm1$ & \textbf{13.7} & \textbf{221.1} & \textbf{33.2}& \textbf{-5.50}& \textbf{174.8} & \textbf{2.99 } & \textbf{$-$} & \textbf{$-$} \\
        & Fastest  & 489 & 102 & 9 & 13.4 & 218.8 & 32.5 & -5.23 & 164.7 & 2.97 & 55 & 8.8\\
        \cmidrule{2-13}
        & \textbf{Sgr} & \textbf{$-$}& \textbf{$-$} & $\sim 4\pm1$ & \textbf{16.9} & \textbf{220.3} &  \textbf{68.1} & \textbf{-6.33}& \textbf{206.7}& \textbf{3.00} & \textbf{$-$} & \textbf{$-$} \\
        & Median & 665 & 102 & 15 & 17.0 & 201.4 & 52.4 & -5.32 & 186.7 &  3.02 & 50 & 12.9 \\ \cmidrule{2-13}
        & \textbf{Sgr} & \textbf{$-$}& $\mathbf{-}$ & $\sim 4\pm1$ & \textbf{19.7} & \textbf{232.9} & \textbf{100.0} &\textbf{-7.04} &\textbf{231.2} &\textbf{3.00} & \textbf{$-$} & \textbf{$-$} \\
        & Slowest & 988  & 102 & 20 & 18.2 & 246.4 & 103.6 & -7.46 & 223.6 & 2.98 & 48 & 16.2
\enddata
\tablecomments{There is variation in the true position of Sgr because we consider orbits drawn from the uncertainties in Sgr's current position to obtain the different kinematics of Sgr. We also list the time ($t_\mathrm{peri}$) and closest Galactocentric approach ($r_\mathrm{peri}$) of Sgr's last pericentric passage for each simulation.}
\label{tbl:SgrPos}
\end{deluxetable*}

To figure out where to place the satellite in our simulation such that it most resembles the Sgr satellite, we have to integrate Sgr's present-day position backward in the potential of our equilibrium simulation. This is not straightforward given that the equilibrium simulation forms a bar and spiral arms after it has been evolved for 3-5 Gyr. There are several different methods for calculating the potential for the simulations. The first is the exact potential calculation that involves direct summation over each particle. However, this can be computationally expensive and not feasible. For that reason, we decide to approximate the potential using the self-consistent field (SCF) basis function expansion described in \citet{Hernquist1992}. We use the \texttt{galpy} Python package\footnote{\url{http://github.com/jobovy/galpy}} \citep{galpy} to calculate the SCF expansion coefficients and initialise the potential using the expansion orders of $N_h=L_h=5$ for the halo, $N_d=L_d=15$ for the disc, and $N_b=N_b=3$ for the bulge. The order of the expansion reflects the complexity of that component. The SCF expansion also requires that we define scale radii for each component. We use the Hernquist radius of the halo, the disk height, and the bulge size defined in \secname \ref{sec:MWmodel} for the halo, disc and bulge respectively. Since the coefficients are additive, we calculate the coefficient on subsets of the particles for each component before adding them together to calculate the approximate potential expansion.  

\begin{figure}
    \centering
    \includegraphics[width=0.45\textwidth]{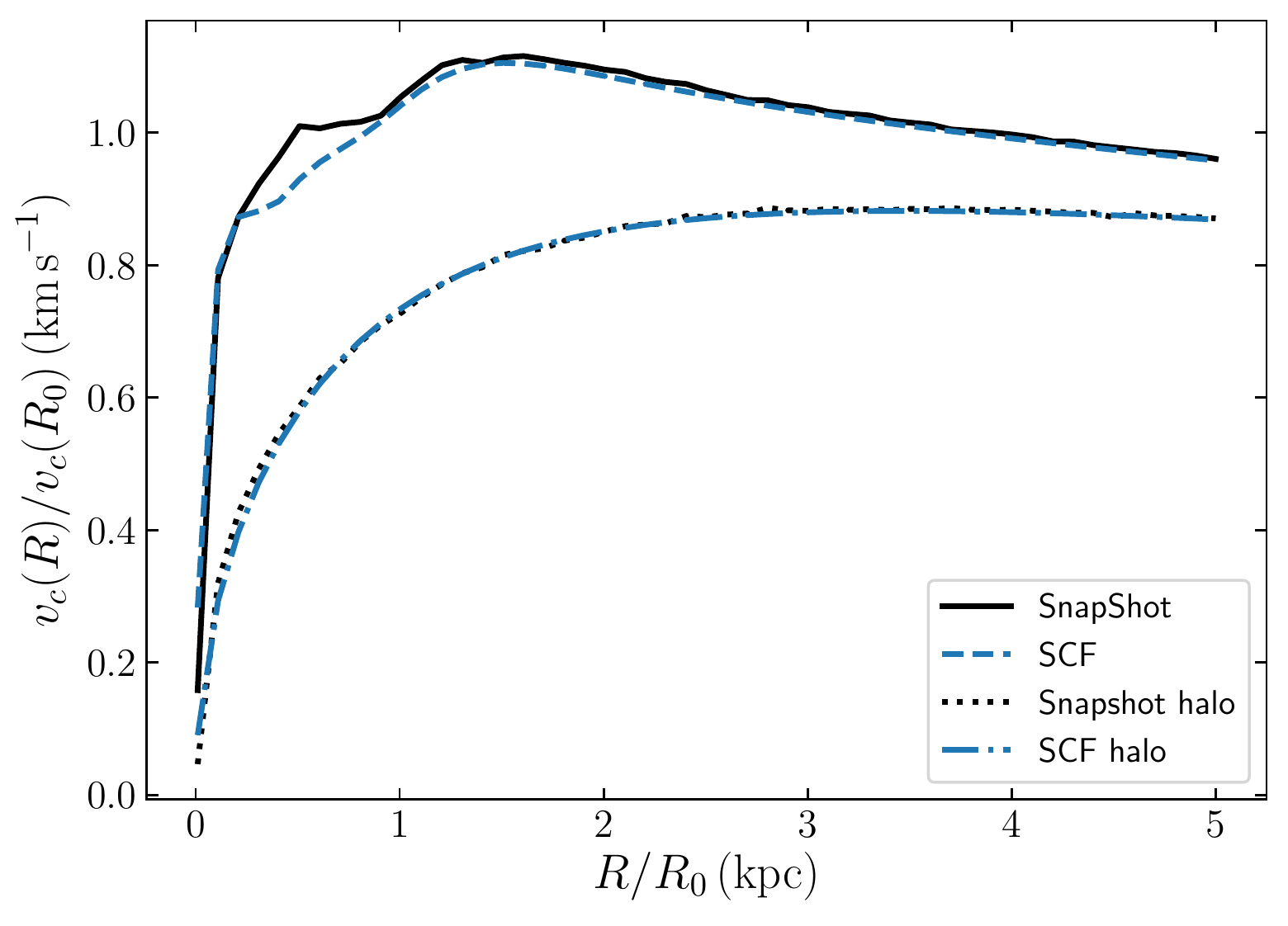}
    \caption{Rotation curve for the full three-component \texttt{galpy} snapshot potential (black, solid) as well as the SCF basis expansion potential (blue, dashed). Also plotted is the rotation curve for just the halo component of the snapshot potential (black, dotted), as well as that of the SCF potential (blue, dash-dotted). While the three-component potential is quite different from the other within 8 kpc, they agree well further out and are fully consistent at all length scales for the halo.}
    \label{fig:MW_rot}
\end{figure}
    
To ensure that the approximated potential behaves similarly to the actual potential of the particles, we test several different properties. The first thing we check is that the rotation curve calculated using each particle's position and mass using direct summation matches the rotation curve of our SCF expansion. We calculate the direct-summation result using \texttt{galpy}'s \texttt{SnapshotRZPotential} that creates an axisymmetrized version of the $N$-body potential by averaging the direct-summation forces at four azimuths spaced 90$^\circ$ apart. \figurename \ref{fig:MW_rot} shows the rotation for both cases as well as the rotation curve of the two halos. We do find that the two disagree within $\sim 8$ kpc, but this is due to the fact that the SCF expansion relies on spherical harmonics as basis functions and it is therefore difficult to reproduce a disc-like structure. In all of our simulations, Sgr does not come within 10 kpc of the centre of the Milky Way, so the halo will be the component that most affects the dynamical friction and orbit of Sgr. It is therefore sufficient that the SCF approximation of the halo matches that of the exact simulation particles for the halo. Though not pictured, we also check the density profile of each component compared to the axisymmetrized density calculated using \texttt{SnapshotRZ} and found that much like the rotation curve, the two were consistent outside $\sim 8$ kpc.

\begin{figure*}
    \centering
    \includegraphics[width=\textwidth]{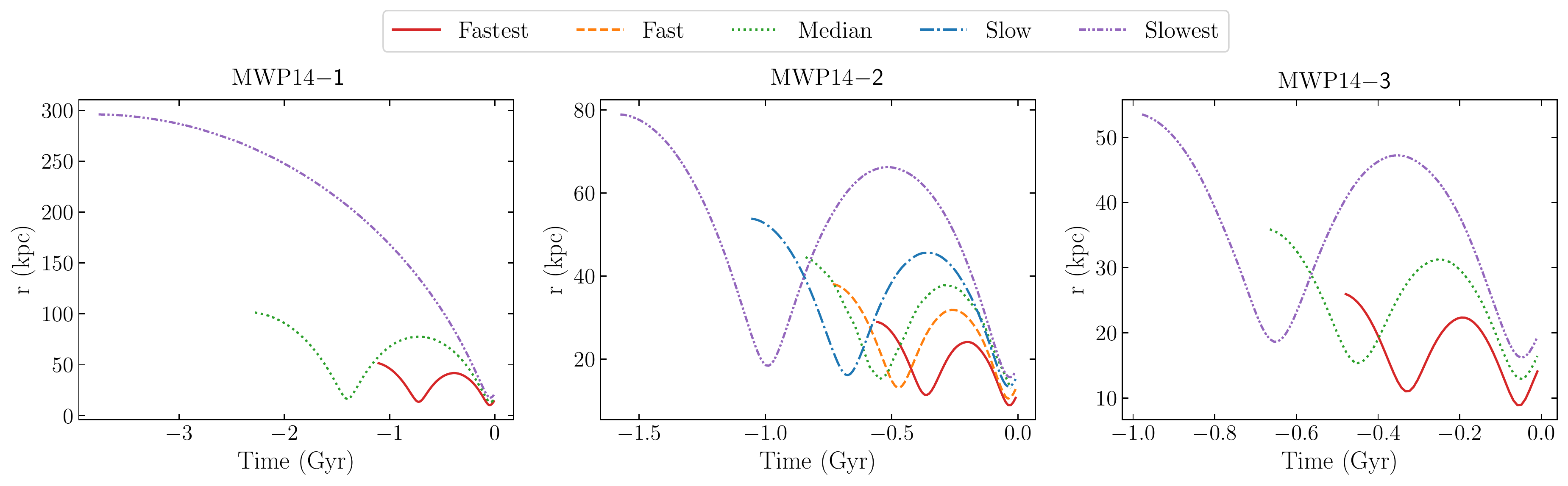}
    \caption{Sgr orbits for the different perturbed simulations. The three different panels show the three different Milky Way potentials. The lines on each panel show the orbits for the different speeds of Sgr: Fastest (red, solid), Fast (orange, dashed), Median (green, dotted), Slow (blue, dash-dotted), and Slowest (purple, dash-dot-dotted). The offset in the different orbits arises from the fact that the different speeds of orbits arise from the uncertainty in Sgr's current position.}
    \label{fig:orbits}
\end{figure*}

We also ran a quick 100 million particle simulation such that we could calculate the approximate the mass of Sgr as a function of time. Using this as a model for mass loss, we were better able to predict the strength of the dynamical friction throughout the orbit. For both the density function required to calculate dynamical friction and the orbiting potential, we used our SCF potential approximation. To calculate the dynamical friction we use the \texttt{ChandrasekharDynamicalFrictionForce} from \texttt{galpy} that calculates the Chandrasekhar dynamical friction on a satellite given the mass. The force from dynamical friction on the satellite is calculated using
\begin{equation}
\begin{split}
    F(x,v) = -2\pi\left[G M\right]&\left[G\rho(x)\right]\ln\left(1+\Lambda^2\right)\\& \times\left[\mathrm{erf}(X)-\frac{2X}{\sqrt{\pi}}\exp{\left(-X^2\right)}\right]
\end{split}\end{equation}
where $G$ is the gravitational constant, $M$ is the mass of the satellite, $x$ is the position of the satellite, $v$ is the velocity of the satellite, and $\rho$ is the background density of the Milky Way. The two calculated parameters are given by $X=|v|/\sqrt{2}\sigma_r(r)$ and 
\begin{equation}
    \Lambda=\frac{r/\gamma}{\max(r_\mathrm{hm},GM/|v|^2)},
\end{equation}
where $r_\mathrm{hm}$ is the half mass radius of the Sgr mass model and $\gamma=\max(|\mathrm{d}\ln\rho/\mathrm{d}\ln r|,1)$. We initialise the dynamical friction object using the half-mass radius and total mass of the combined dark matter and stellar component of each Sgr model. For the potential in which the satellite is moving we use the SCF potential calculated for the equilibrium initial condition. Next, as we integrate the satellite backward, we update the mass of Sgr after each of 1000 equally spaced integrals in the total integration time using the results of the 100 million particle simulation. This is only a rough estimate of mass loss for two reasons. First, the mass loss is going to change between simulations as the orbit and Milky Way density changes, so using the mass loss from our test simulation will not be exactly accurate. Second, we do not change the half mass radius for each interval, which means the Sgr will be approximated as too diffuse. However, we find that despite these assumptions, our placement of Sgr results in a fairly accurate final position in our simulations. 

Our initial conditions for Sgr are taken from the analysis of Sgr's orbit considered in BB21. Within this exploration of the orbit, they initialised over 10,000 orbits where the initial conditions are drawn from Gaussian distributions given the errors in Sgr's position and velocity. The position was given by $R=26\pm2$ kpc \citep{McConnachie2012} and $\{\alpha,\delta\}=\{283.8313, -30.5453\}$ deg \citep{GaiaSatKin}. The
velocities and associated uncertainties were $v_\mathrm{los} = 140\pm2$ km\,s$^{-1}$ \citep{McConnachie2012} and   $\{\mu_\alpha^*,\mu_\delta\} = \{-2.692, -1.359\}\pm0.001\,\mathrm{mas\,yr}^{-1}$ \citep{GaiaSatKin}. We chose $X_\mathrm{sun}=8.1$ kpc for consistency with \texttt{galpy} that we use for our orbit integration. 

In each section of our analysis, we have a more thorough discussion of how we choose the final position of Sgr in each simulation because it depends on the Milky Way potential, the Sgr model, and the Sgr orbital kinematics. Once we know where we want Sgr to end, we integrate the orbit back two apocentres. We do this for several reasons. First, we go back to an apocentre because the modeling assumption that the disc is in equilibrium before the interaction is best satisfied when Sgr is at an apocentre so it is perturbing the disc as little as possible to start. Second, we go back two apocentres because we want to go back far enough to get the secondary effects from the passage, but longer than two apocentres, and the backward orbit integration becomes more uncertain. Furthermore, in BB21 both the model and one-dimensional self-gravitating $N$-body simulations showed that the 2 apocentres was sufficient to capture the overall trend in the asymmetry and mean vertical velocity. 

Once we have the position and velocity of Sgr's orbit two apocentres ago, we use it to place our Sgr particles in our simulation and truncate all particles at 500 kpc. While we use this method to try and estimate where the satellite will end up later in the simulation, this does not guarantee it will finish where we expect.  

\subsection{Sgr properties throughout the simulation}

Throughout the simulation, we track the Sagittarius particles to obtain the orbit and effective mass of the satellite. This is important if we want to compare the asymmetry in the simulation to the model developed in BB21. To do this, we use \texttt{clustertools}\footnote{\url{https://github.com/webbjj/clustertools}} \citep{webb2020}, a \textsc{Python} package developed to retrieve the properties of clusters in simulations. We start by initializing a cluster using the positions, velocities, and masses of both the stellar and dark matter particles in our satellite. We can then use \texttt{clustertools} to find the central density of our cluster, the mass within the tidal radius, and the half-mass radius. 

When setting up our simulation, we did our best to place the satellite such that it would end up around Sgr's current position. However, we could not predict its path exactly, so once the simulation had finished, we chose the snapshot that most closely resembled the location and velocity of Sgr. \figurename \ref{fig:orbits} shows the orbit of Sgr extracted from each of our simulations and provides an approximate timescale for each. Table \ref{tbl:SgrPos} shows the position of Sgr drawn from the distributions in the position and velocity arising from the errors in the current phase-space parameters. We also include the closest position of our satellite in the different simulation scenarios. For the three different Sgr mass simulations, the benchmark position does not change. However, by definition, when we look at how the uncertainty in Sgr's current day position affects the orbit and therefore the perturbation, the benchmark changes. In some cases, we see the mass increase from one time step to another. This is because we define the mass as the total mass from Sgr particles within the tidal radius at each time step. Since Sgr has just passed through pericentre, some of the stream particles catch up to Sgr from one time step to another therefore artificially increasing the observed mass of Sgr. Finally, we also give the time ($t_\mathrm{peri}$) and Galactocentric distance ($r_\mathrm{peri}$) of Sgr's last pericentric passage for each simulation in Table \ref{tbl:SgrPos}. As is clear from \figurename \ref{fig:orbits}, the time of the last pericentric passage is $\approx 50\,\mathrm{Myr}$ with only a few tens of Myr variation between different simulations---much less than the dynamical timescale of Sgr's orbit. 

\section{Investigating the Solar Neighbourhood}\label{sec:solarneighbourhood}

In this section, we investigate how both the mass of Sgr and its kinematic properties affect the vertical dynamics in the solar neighbourhood. As in BB21, we focus on the effect on the number density $n(z)$, characterized through the asymmetry $A$, and on the mean vertical velocity. The asymmetry is defined as
\begin{equation}\label{eq-asymmetry}
    A(z) = {n(z) - n(-z) \over n(z) + n(-z)}\,,
\end{equation}
where $z$ is the vertical position with respect to the mid-plane. The density, which is used to calculate the asymmetry, and the mean vertical velocity are the zeroth and first velocity moments of disturbed phase-space spirals. For this reason, we will also look at the phase-space spiral.

Before placing Sgr, the solar neighbourhood could exist at any $\phi$ value in the simulation. However, by placing Sgr, the symmetry is broken and the solar neighbourhood is now at $R=8.1$ kpc and $\phi=0$ when $t=\mathrm{now}$. \tablename \ref{tbl:SgrPos} shows the final positions of Sgr in our simulations alongside the true position of Sgr in the Milky Way. For each simulation we find the two nearest snapshots because our orbits in the SCF potential, which overlooks rotation and reflex motion of the Milky Way, are not able to exactly predict the orbit and therefore none of the snapshots are exactly correct. Once we have the closest snapshots, we choose the snapshot that has the closest $z$ position and $v_z$ velocity to the true Sgr position using the following equation: 
\begin{equation}
    \Delta= \sqrt{\left(\frac{z_{\mathrm{sim}}-z_{\mathrm{true}}}{8\,\mathrm{kpc}}\right)^2+\left(\frac{v_{z,\mathrm{sim}}-v_{z,\mathrm{true}}}{220\,\mathrm{km\,s^{-1}}}\right)^2}
\end{equation}
where the subscript `sim' denotes the simulation's position and velocity and `true' denotes the position where we placed Sgr when integrating backward. We also looked at the asymmetry and mean vertical velocity for both snapshots. We found that they were very similar considering the uncertainties, and therefore it was not worth plotting both.

To select the solar neighbourhood, we adjust for the centre of mass and angular momentum of the disc, such that the disc is centered on our coordinates and is aligned with the plane $z=0$. For each solar neighbourhood-like volume, we chose a cylinder centered at $\{X,Y,Z\}=\{8.1,0,0\}$ kpc with a radius of 1 kpc. Throughout this section, we compare our simulations to the perturbation measurements from \citet{Bennett2019} who use \Gaia DR2 data within a cylinder radius of 250 pc. To ensure consistency, we did verify that the asymmetry and the mean vertical velocity measurements of the perturbation were consistent when comparing volumes of 250 pc and 1 kpc in our simulation. We chose to use 1 kpc however, because of the much larger number of particles and subsequent decrease in the uncertainty in our measurements. This allows us to choose the location similar to the solar neighbourhood in distance from the centre of the disc. To properly account for any warping for of the disc, once we have selected the volume of interest in our simulation, we adjust for the angular momentum of the volume. To do this, we select a volume centred on the same location and with a radius twice as large. Adjusting for the angular momentum of this larger volume means that when we reduce the radius of the cylinder to the radius specified, we capture all of the particles that are shifted into our volume by the adjustments that would have been missed if we had only looked at the specified volume. 

Another test of our models is how well the final remnant mass compares to the current estimates of Sgr's thus far. \citet{Law2010} used the velocity dispersion of the stellar tidal stream to estimate a currently bound mass of $2.5_{-1.0}^{+1.3}\times10^8\,\,\mathrm{M_\odot}$. In \citet{Frinchaboy2012}, they found a half-mass radius of Sgr of $r_\mathrm{hm}=1.2\times10^8\,\,\mathrm{M_\odot}$ that translates to a mass of $\sim4\times10^8\,\,\mathrm{M_\odot}$ within 5 kpc. Finally and most recently \citet{Vasiliev2020} concluded that Sgr has a total mass (including dark matter) of $(4\pm1)\times10^8]\,\,\mathrm{M_\odot}$. Though these are all estimates of the true remnant mass and there are not yet any direct measurements, they are all of a similar range. 

Finally, we use the model developed in BB21 as a tool to interpret the results of our simulations. In BB21, the model was thoroughly tested against self-gravitating one-dimensional simulations. While fitting the model to the results of our simulations will be left to future work, we find that it is useful for understanding the trends seen in the different simulations.

\begin{figure}
    \centering
    \includegraphics[width=0.45\textwidth]{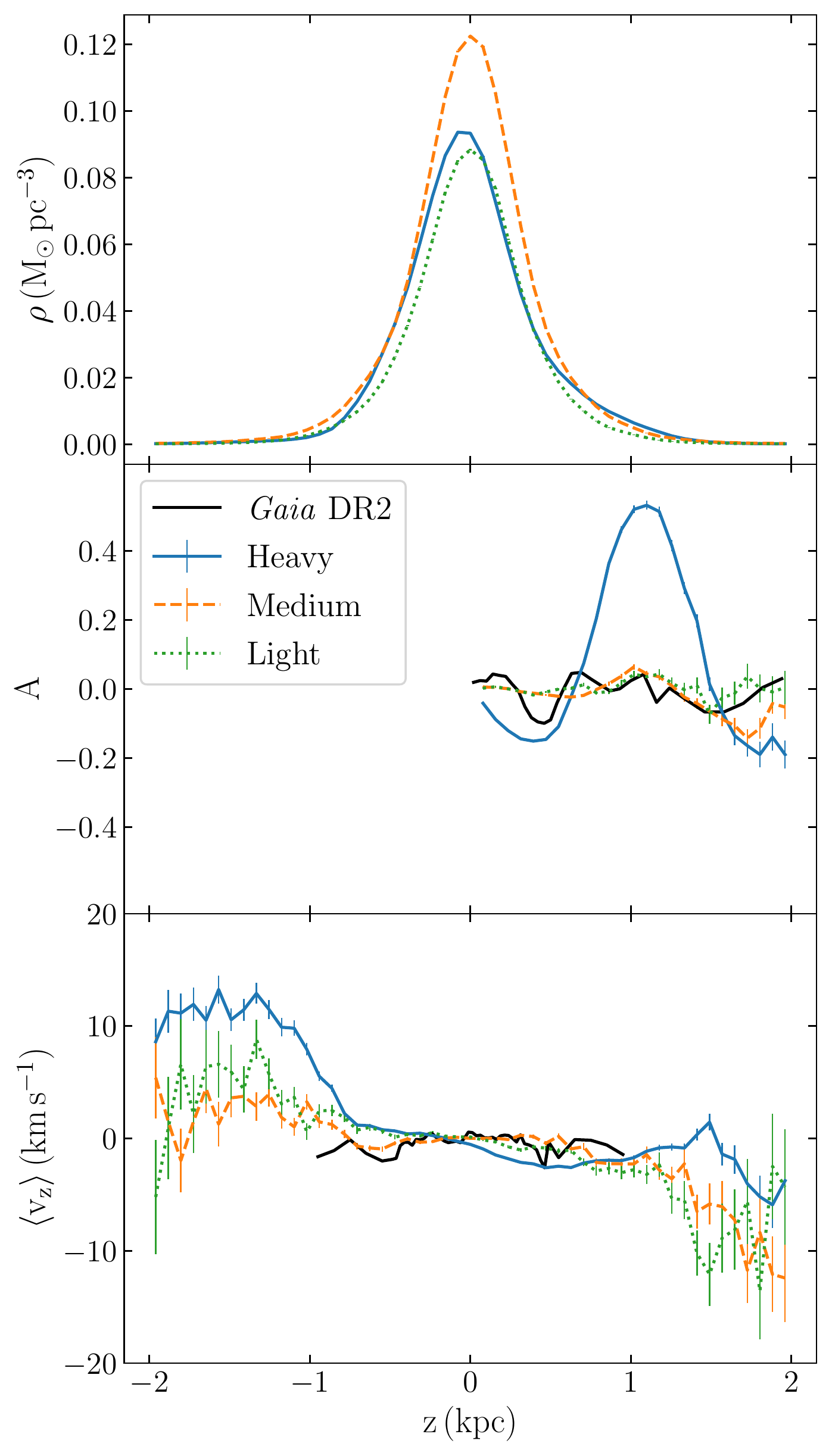}
    \caption{\textit{Top:} Vertical density of solar neighbourhood-like sections of the simulation in cylinders with a radius of 1 kpc at $R=8.1$ kpc and $\phi=0$ for three different Sgr mass models. From heaviest to lightest Sgr mass model, the Sgr models in each simulation are Heavy (blue solid), Medium (orange dashed), and Light (green dotted). 
    \textit{Middle:} Vertical number count asymmetry for the same volumes and simulation as the top panel. Only $z>0$ is plotted as the asymmetry is antisymmetric by definition. The black line shows the Gaia asymmetry from \citet{Bennett2019}.
    \textit{Bottom:} Mean vertical velocity for the same volumes and simulations. 
    The \midplane densities are all near the true value of the Milky Way mid-plane density ($0.1\,\mathrm{M_\odot\, pc^{-3}}$). None of the asymmetries are consistent with the Gaia data in shape or in amplitude. The mean vertical velocities do not achieve any true signal within the range of the data to determine a definitive match or mismatch.}
    \label{fig:Asym_mass}
\end{figure}

\subsection{Changing Sgr mass}\label{sec:mass}

\begin{figure}\hspace{-0.5cm}
    \includegraphics[width=0.52\textwidth]{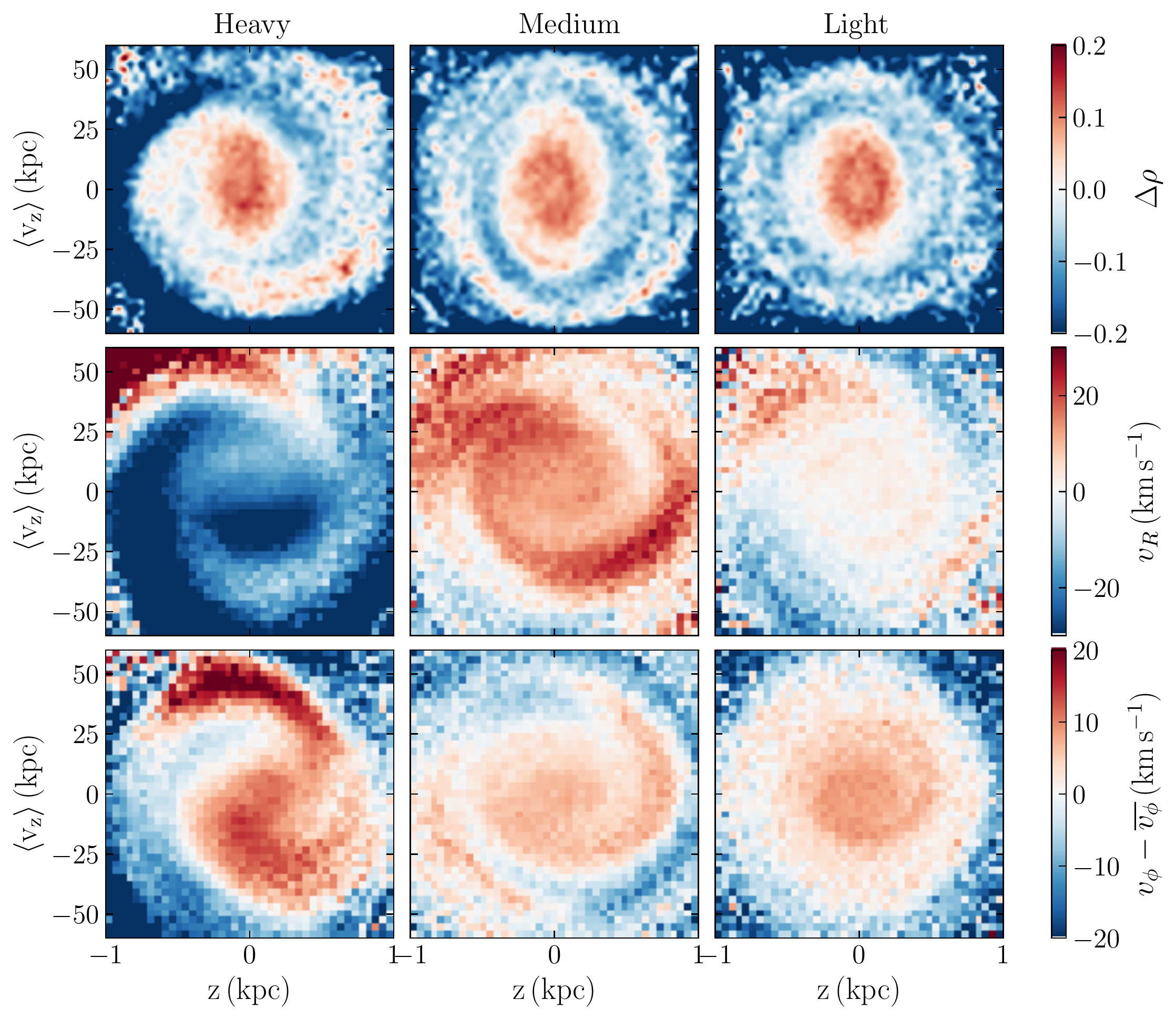}
    \caption{Phase-space density for three simulations with varying Sgr mass models from heaviest on the left to lightest on the right. The volume considered is a cylinder centered at $R=8.1$ kpc with a radius of $r_\mathrm{cyl} = 1$ kpc. The top row shows the spiral coloured by $\Delta\rho$, the middle row is coloured by the radial velocity, and the bottom row is coloured by the normalized azimuthal velocity. The phase-space spiral is more loosely wound than what is seen in the \emph{Gaia} DR2 data.}
    \label{fig:phasespace_mass}
\end{figure}

We start by looking at the effect of changing Sgr's mass on the vertical perturbations to the disc. While this might seem like a simple scaling problem at first, changing Sgr's mass will also affect the dynamical friction and therefore the orbit of Sgr, complicating matters. In our simulations, we choose Sgr mass models Sgr 1, Sgr 2, and Sgr 3 from BB21 as described in \secname \ref{sec:Sgrmodel}, referred to as Heavy, Medium, and Light models, respectively, throughout the rest of the paper.  These satellites orbit in MWP14-2 from BB21 as described in \secname \ref{sec:MWmodel}. We do not investigate the other two models discussed in BB21, Sgr 4 and Sgr 5, as the amplitude from Sgr 3 is already too small to match the observed perturbation, so a smaller Sgr model certainly would not reproduce the perturbation either. As previously discussed, we look at a volume of a cylinder with a radius of 1 kpc in each of the simulations. In the Heavy simulation there are 1\,858\,959 particles, the Medium simulation has 2\,310\,317 particles and the Light simulation has 1\,632\,302 particles.  

We start by looking at the density profile of the disc, the asymmetry and the mean vertical velocity as a function of height above and below the disc. \figurename \ref{fig:Asym_mass} shows the density in the top panel. While the Heavy and Light simulation have similar mid-plane densities, the mid-plane density of the Medium simulation is quite a bit larger. This larger mid-plane density is caused by differences in the location of the spiral structure that develops in our equilibrium simulations and that is exacerbated by the influence of Sgr. The larger mid-plane density of the disc means that the effective mass of the disc at the solar neighbourhood is larger. With a larger mass, the disc is more difficult to perturb and we would therefore expect a smaller relative amplitude in the perturbation compared to the other two simulations (assuming all other things are equivalent which is not the case). The effects of the mid-plane density on the asymmetry was investigated in BB21, which confirmed that a larger mid-plane led to a smaller response of the disc. 

The middle panel of \figurename \ref{fig:Asym_mass} shows the asymmetry for each of the Sgr models in our simulation at the solar neighbourhood. It is clear that the Heavy Sgr mass model resulted in the largest perturbation amplitude. However, we find that the amplitudes of the Medium and Light Sgr mass models are similar. This is unexpected as a larger mass means a larger force and should therefore lead to a larger perturbation. It is likely that this discrepancy is explained by the larger mid-plane density of the disc in the Medium simulation. In terms of the period of the oscillation, we see that the wavelength of the asymmetry decreases as we decrease the mass of Sgr, though this is difficult to confirm from only the three simulations. Since the change in the mid-plane density affects the mass of the disc, it will also result in a change to the vertical frequency of the disc, and could therefore affect the asymmetry wavelength. Finally, the black lines in \figurename \ref{fig:Asym_mass} show the the measured asymmetry from \emph{Gaia} DR2 taken from \citet{Bennett2019}. Clearly, none of the simulations match the shape, wavelength, or amplitude of the data, especially at $z\sim0.4$ kpc. While the discrepancy in the amplitude and possibly even the wavelength, can be explained by changes in the mid-plane density, the shape of the asymmetry cannot be as easily fine-tuned to create a match and it is unlikely that changes to the disc would affect the perturbation to the degree it needs to be altered to force this match. 

The last panel in \figurename \ref{fig:Asym_mass} shows the mean vertical velocity as a function of height in the three different simulations. All three models appear to display a breathing mode near the mid-plane of the disc, particularly the Heavy and Light models. Like the asymmetry, the amplitude of the perturbation to the mean vertical velocity in the Medium simulation is likely suppressed by the higher mid-plane density. Unfortunately, the mean vertical velocity measurement form \emph{Gaia} DR2 only goes out to $\pm$1 kpc and at that distance the amplitudes of the simulation measurements are quite small. However, while the Medium model cannot be ruled out using the mean vertical velocity, the Heavy model is inconsistent with the data at all heights, and the Light model does a poor job of matching the \Gaia data at negative heights. Looking at both the asymmetry and the mean vertical velocity leads us to conclude that it is unlikely any of these three Sgr models could have led to the observed oscillation in the Milky Way.

The asymmetry and mean vertical velocity are both derived from the phase-space density. \figurename \ref{fig:phasespace_mass} shows the phase-space spiral for the three simulations coloured by the normalized density change, $\Delta\rho$, the radial velocity, and the normalized median subtracted azimuthal velocity. Looking first at the density phase-space spiral, comparing the different simulations, it is difficult to uncover trends in the tightness of the spiral. However, it is evident that the Heavy simulation has a larger perturbation than the other two, especially on the edges of the spiral. Comparing the simulations to observations, we see that the spirals in the simulation are much less tightly wound than the \Gaia DR2 data \citep{Laporte2019}. By 1 kpc, the spiral in \emph{Gaia} DR2 has wound around 2-3 times, whereas our simulation has wound only once. Next, we look at the phase-space spiral coloured by the radial velocity. It is very evident that as we decrease the mass of Sgr, the absolute amplitude of the radial velocity decreases. This is an interesting trend, as it does not appear to have been affected by the different mid-plane densities like our other measurements. It also appears that the spiral becomes more tightly wound as we decrease the mass of Sgr. Finally, when comparing the different phase-space volumes coloured by the normalized azimuthal velocity, we see very distinctly that the spiral becomes more tightly wound as we decrease the Sgr mass. 

After looking at how changing Sgr mass affects the perturbation to the solar neighbourhood, it is safe to conclude that the median velocity orbit for any of the models will not result in a match to the \emph{Gaia} DR2 data. Our next step is to investigate whether or not changing the kinematics of Sgr will help achieve a match.

\subsection{Changing Sgr kinematics}\label{sec:SgrVel}

\begin{figure}
    \centering
    \includegraphics[width=0.45\textwidth]{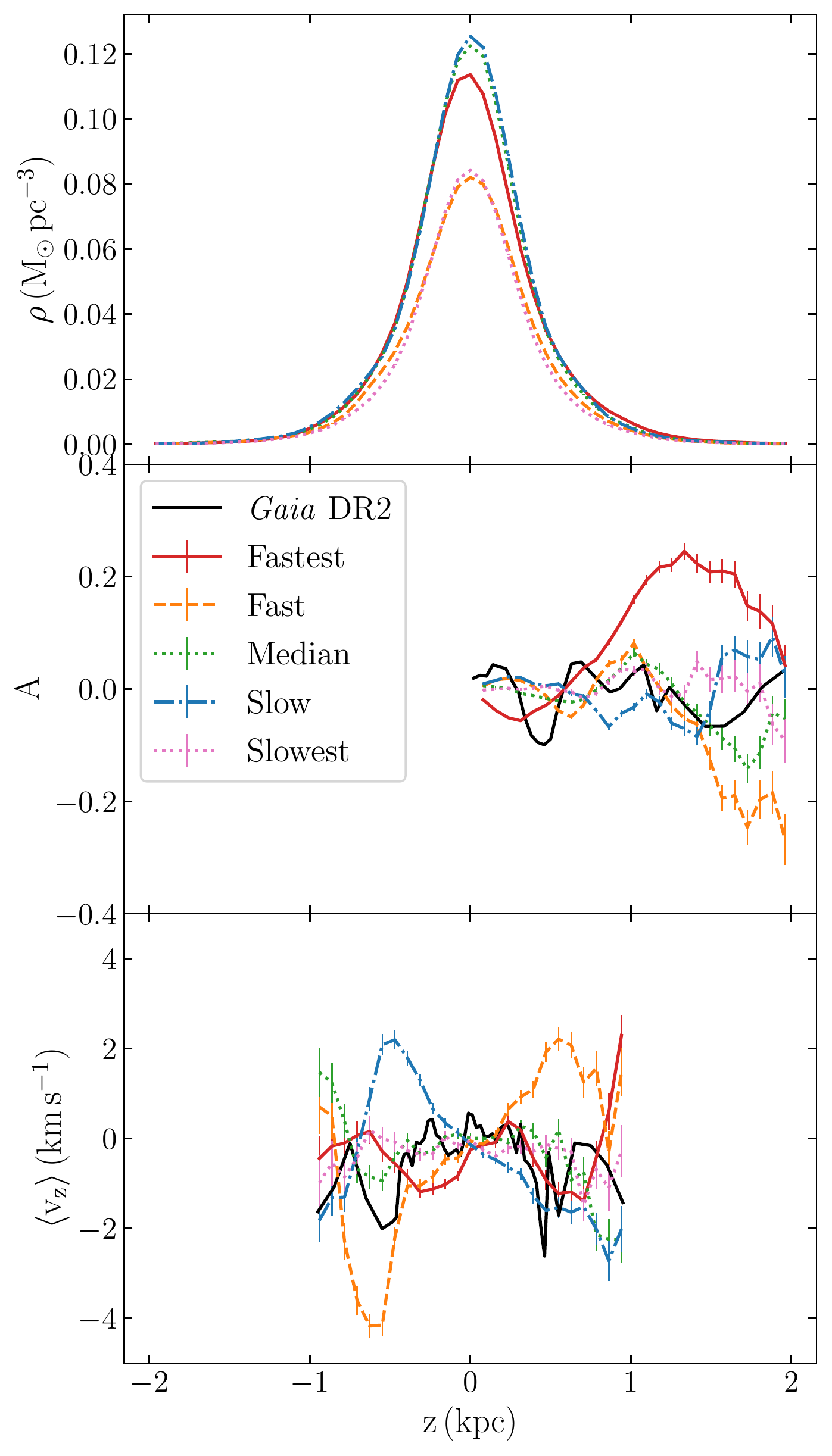}
    \caption{\textit{Top:} Stellar density as a function of height in a solar neighbourhood-like location. The volume considered is vertical cylinder with a radius of 1 kpc at a position of $R=8.1$ kpc and $\phi=0$. The four coloured line indicate the four different Sgr orbit models: Fastest (red solid), Fast (orange dash), Slow (green dot) and Slowest (blue dash-dot).
    \textit{Middle:} Vertical number count asymmetry for four different Sgr velocity simulations. Only $z>0$ is plotted as the asymmetry is antisymmetric by definition. The black line shows the Gaia asymmetry from \citet{Bennett2019}.
    \textit{Bottom:} Mean vertical velocity in the solar-neighbourhood-like volume. The uncertainties in the simulation outside $|z|<1$ were larger than the signal. Since we only have the \emph{Gaia} data within the same heights, we choose to omit the simulation values outside that area as well.
    The \midplane densities vary between ($(0.08-0.13)\,\mathrm{M_\odot\, pc^{-3}}$), which spans the true value of the Milky Way mid-plane density. None of the asymmetries or the mean vertical velocities match the \emph{Gaia} DR2 observations. }
    \label{fig:vel_asym}
\end{figure}

\begin{figure*}
    \centering
    \includegraphics[width=0.95\textwidth]{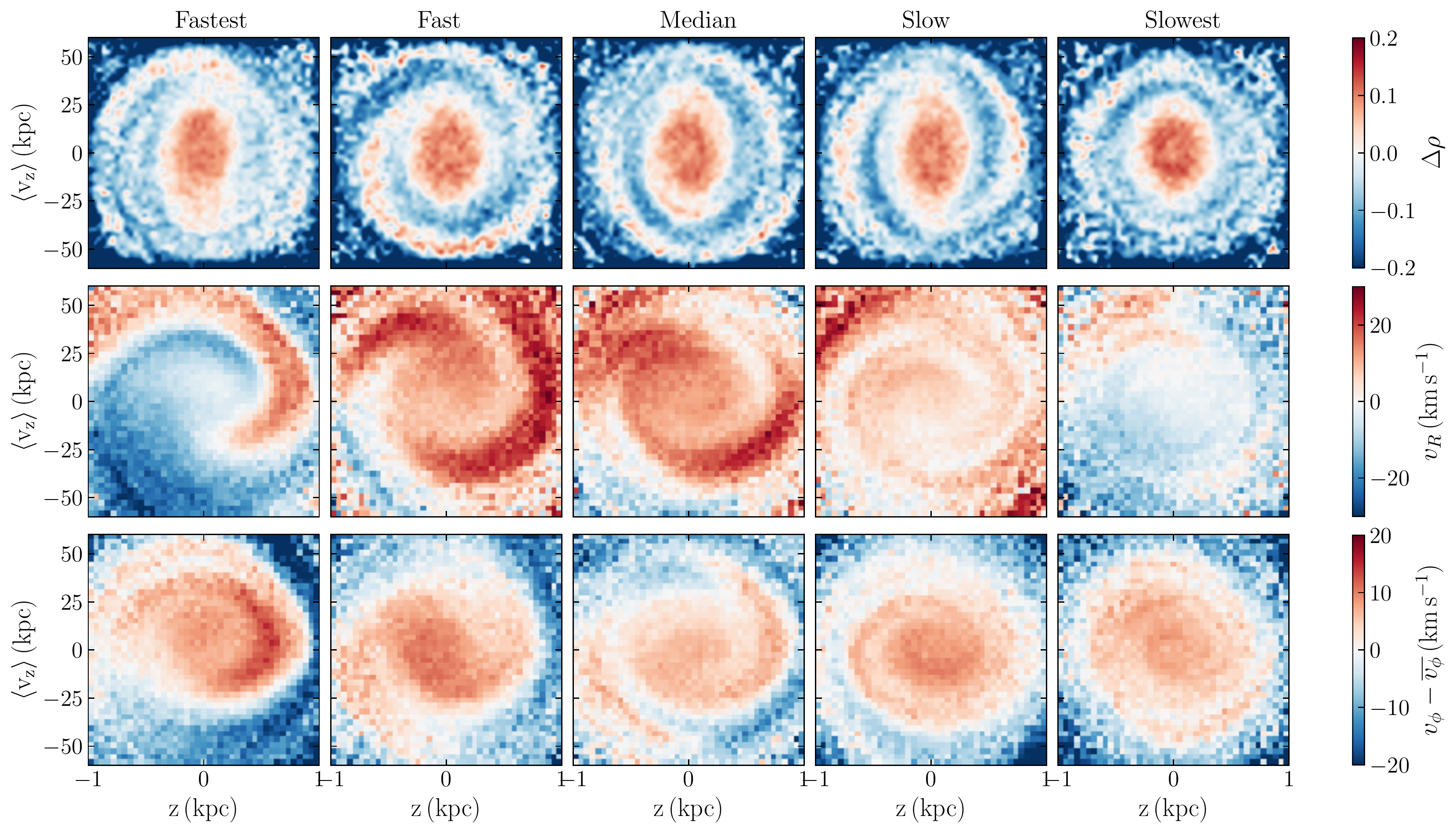}
    \caption{Phase-space spiral for five different Sgr orbits, parameterized by the speed at which Sgr passed through the disc most recently from Fastest on the left to Slowest on the right. The different rows correspond to the normalized change in density (top), the radial velocity (middle), and the normalized azimuthal velocity (bottom).}
    \label{fig:phase_vel}
\end{figure*}

Uncertainty in Sgr's current position and velocity, as well as uncertainty in some properties of the Milky Way such as the mass of the halo, leads to fairly large uncertainty in the orbit of Sgr. One method of parameterizing the different possible orbits of Sgr is the speed at which Sgr was travelling when it passed through the \midplane of the disc, $v_{z,\mathrm{through}}$, and how long ago that occurred, $t_{\mathrm{through}}$. These two parameters are highly correlated for the current uncertainties in Sgr's present phase-space position and in the Galactic potential, so looking at different velocities of Sgr corresponds to also looking at different times since passing through the disc. For our investigation of the kinematics of Sgr, we chose to use the Medium mass model of Sgr, because the asymmetry from the Heavy simulation had an amplitude that exceeded observations and the amplitude from the Light model was very small. It is worth noting that in our investigation into the effects of the mass of Sgr, even the lightest model finished with a mass that was much larger than the estimated mass of the Sgr remnant today. Therefore, by choosing the Medium Sgr mass model, in the cases where we find that the amplitudes of the asymmetry is still not large enough, we are also able to rule out the Light Sgr mass model as suggested by \secname \ref{sec:mass}.

BB21 looked at Sgr’s orbit as a function of the uncertainties in its current position by sampling 10,010 initial conditions from the uncertainty distributions in the current positions and velocities of Sgr \citep{GaiaSatKin}. Using these orbits, we calculated the most recent velocity through the \midplane and the time since passing through the \midplane for set of initial conditions. For the purpose of this paper, we bin the orbits by their velocity through the \midplane into nine bins and run simulations for the first (fastest), third (fast), fifth (median), seventh (slow), and ninth (slowest) bin. This allows us to cover a the entire range of Sgr velocities with our simulations. From fastest to slowest, the numbers of particles in each solar neighbourhood was: 2\,239\,746, 1\,662\,069, 2\,310\,317, 2\,360\,833, and 1\,582\,787. The most recent velocity through the \midplane and the time since passing through the \midplane are very strongly correlated with the time and Galactocentric distance of Sgr last pericentric passage given in Table~\ref{tbl:SgrPos}.

\figurename \ref{fig:vel_asym} shows the vertical density distribution, vertical number count asymmetry and mean vertical velocity for our four different Sgr velocity simulations. In BB21 as well as \secname \ref{sec:mass}, we found that changing the \midplane density did not affect the shape of the asymmetry, but it did affect the amplitude as well as the perturbation wavelength. For this reason, when looking at trends in our simulations, we will focus on comparing the Fast and Slowest simulations, and the Median and Slow simulations.

\begin{figure*}
    \centering
    \includegraphics[width=0.8\textwidth]{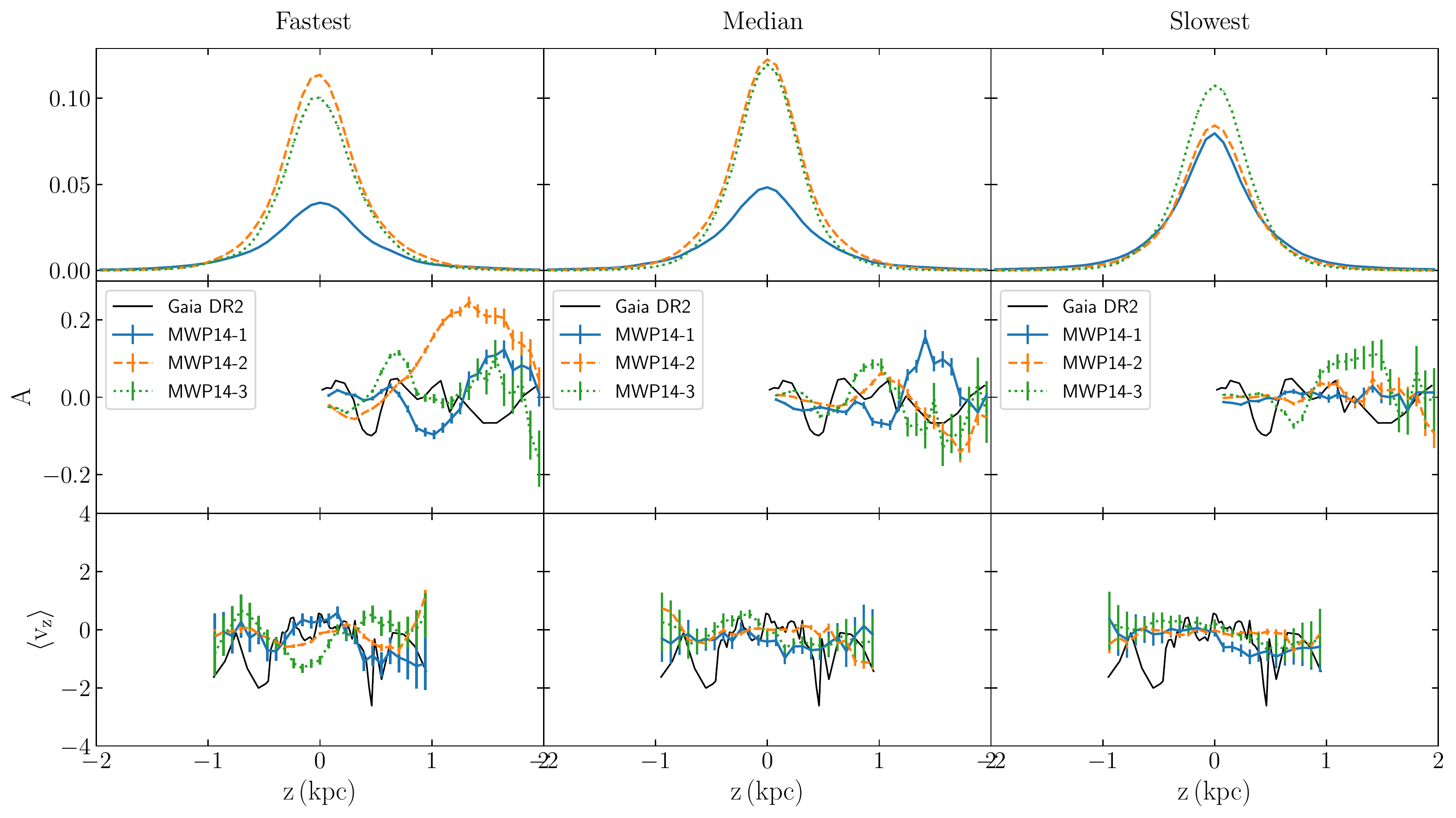}
    \caption{\textit{Top:} Number count asymmetry of the three different Milky Way halo simulations for three different kinematics of Sgr from Fastest on the left to Slowest on the right. The three different Milky Way halos include the lightest MWP14-1 (blue solid), the medium halo MWP14-2 (orange dashed), and the heaviest halo MWP14-3 (green dotted). For MWP14-3, we have only plotted the asymmetry out to $z=1.3$ because the uncertainties become quite large further out and obscure the shape of the asymmetries. 
    \textit{Bottom:} Mean vertical velocity for each grouping of Sgr kinematics and Milky Way halo mass. To match the data, we limit the calculations to $|z|<1$ kpc.
    For both, the observed \Gaia DR2 observations are plotted in black. The closest match to observations is the median Sgr orbit for MWP14-3, though it is still far from a perfect match.}
    \label{fig:MWmass_asym_meanV}
\end{figure*}

The second panel of \figurename \ref{fig:vel_asym} shows the asymmetry for the four different velocity simulations. The Fastest simulation has an amplitude much larger than the four other simulations and does not appear to follow any clear trends when looking at the asymmetry alone. When comparing the Median and Slow simulation, there is no discernible difference in the amplitude. Looking at the asymmetry wavelength, the Median Sgr velocity simulation has a larger wavelength with a peak at $z\sim 1$ kpc while the Slow simulation has peaks at $z\sim 0.3, 1.2, 1.8$ kpc. Finally, we compare the Fast and Slowest Sgr simulations. We see that both the amplitude and the asymmetry wavelength of the Fast simulation is larger than in the Slowest simulation. The Fast simulation has peaks at $z\sim 0.2, 1.1$ kpc while the Slowest simulation has peaks at $z\sim0.4,1.1,1.6$. This agrees with the trend seen in the Median and Slow simulation. This suggests that as Sgr passes through the disc more quickly, the wavelength of the perturbation increases. This supports the trends seen in BB21, where both the amplitude and wavelength decrease as the speed of Sgr decreases. The most prominent feature of the asymmetry of the observed asymmetry is the dip at $z\sim0.4$ that has a fairly large amplitude. None of our simulations match the position of the dip. The Fast simulation is the closest with a dip at $z\sim 0.6$ kpc, but the amplitude is half that of the observed dip and yet from \tablename \ref{tbl:SgrPos}, we see that the final mass of the simulated Sgr is still $\sim3-5$ times heavier than the current estimates of the Sgr remnant, so if we were to increase the mass of Sgr to match the amplitude, we would be greatly overestimating the mass of Sgr.

It is also important to consider how the mean vertical velocity in the asymmetry compares with the true signal in \emph{Gaia} DR2. In the data, \citet{Bennett2019} found a bending mode in the solar neighbourhood within $|z|<1$. This was later confirmed by \citet{carrillo19} who found a bending mode within $|z|<1$ and a breathing mode further out. Both the Fast and Slow simulation show strong breathing modes that are inconsistent with the data. The Slowest simulation has no notable breathing or bending signal within the uncertainty in the points. The Fastest and Median simulation both have a small bending signal, but the amplitude is far off from the true amplitude from \emph{Gaia} and the asymmetry associated with both the Fastest and Median simulations is far from the observed asymmetry. 

Finally, we look at the phase-space spiral of the five different simulations. \figurename \ref{fig:phase_vel} shows the phase-space spiral for all five simulations. We include the phase space coloured by the normalized change in density, the radial velocity, and the normalized azimuthal velocity. As we look from the Fastest simulation to the Slowest, especially once coloured by the normalized azimuthal velocity, we see spiral becomes more tightly wound. This supports the trend seen in our asymmetry where the asymmetry wavelength decreases as Sgr slows down. Though harder to see in the normalized density, in both the radial velocity and normalized azimuthal velocity, we see the absolute amplitude of the of the perturbation approximately decrease as Sgr passes through the disc more slowly. Much like in the simulations where Sgr's mass is changed, we see that the phase-space spiral in our Sgr velocity simulations is much more loosely wound than what has been observed using the \Gaia data. 

Looking at the asymmetry, mean vertical velocity and phase-space spiral, we once again find that none of our simulations are able to reproduce the perturbation seen in the solar neighbourhood. We can therefore conclude that changing the velocity of Sgr is not enough to make the simulated perturbation match the observed perturbation. This is even more obvious when we note that the simulations used in this \secname are run using the Medium Sgr mass model. In \secname \ref{sec:mass}, we established that the Medium mass model was still too large in terms of the remnant mass. So not only is the perturbation not a match for any of the simulations, but the second heaviest Sgr model is still not enough to create the observed amplitude in the asymmetry and yet the final mass is still much too large.

\subsection{Changing Milky Way Halo Mass}\label{sec:MWmass}

\begin{figure}
    \centering
    \includegraphics[width=0.5\textwidth]{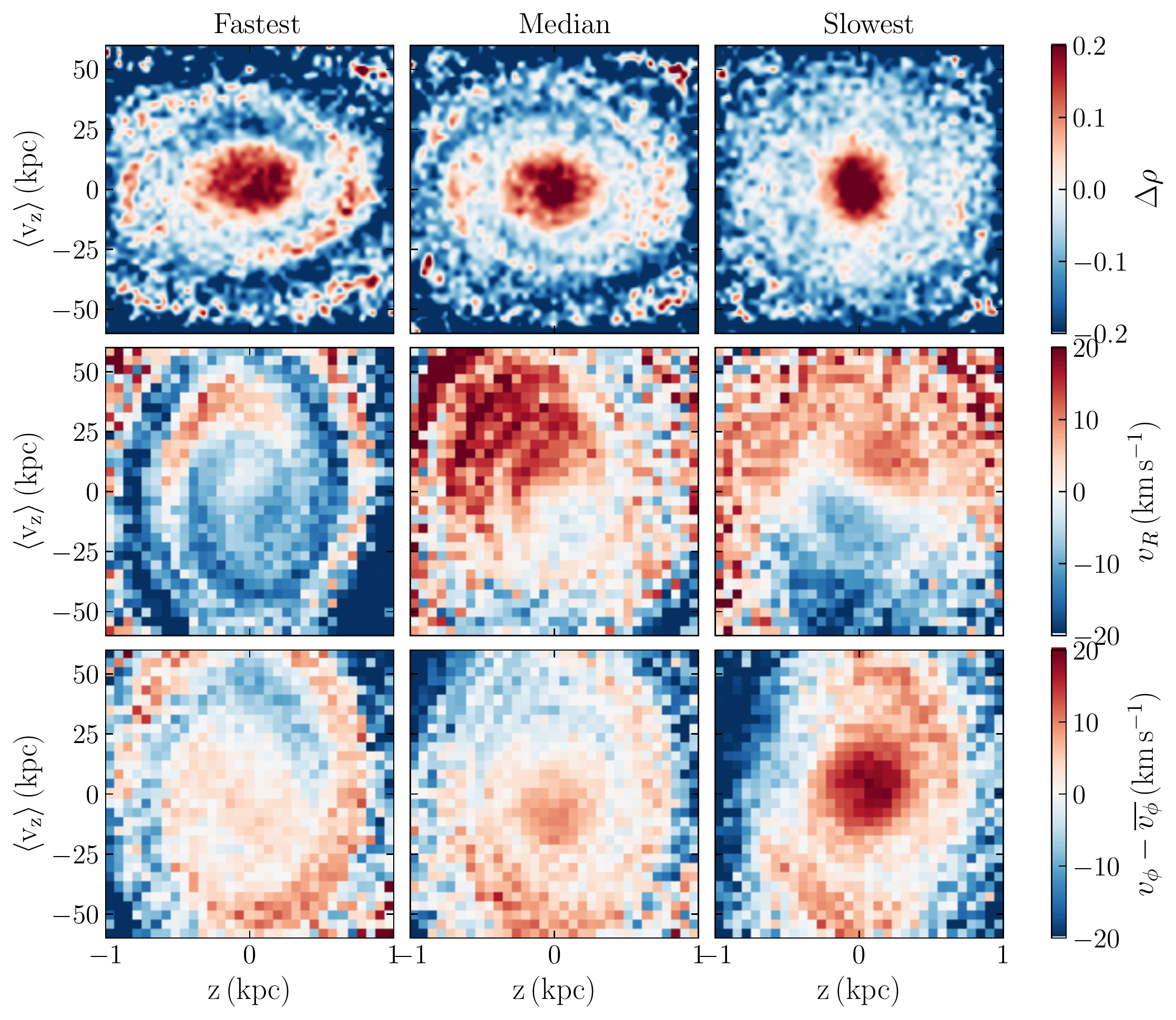}
    \caption{Phase-space spiral for the MWP14-1 simulations,  which is the initial condition with the lightest of the three Milky Way halos. The different columns represent three different speeds of Sgr as it passed through the mid-plane most recently from Fastest on the left to Slowest on the right. The rows represent the different ways of colouring phase space: the normalized change in the density (top), the radial velocity (middle), and the normalized azimuthal velocity (bottom).}
    \label{fig:phase_MW1}
\end{figure}

Our final investigation looks at how changing the mass of the Milky Way changes the perturbation in the simulation. To do this, we use the medium Sgr mass model (Sgr 2). Though we have found that the final mass of the medium Sgr is consistently too large, we want to ensure that the signal is large enough to see in our simulations. We also look at the fastest, median, and slowest Sgr velocities for each of the Milky Way models. Again, we bin orbits by their velocity through the mid-plane ($v_{z,\mathrm{through}}$) for each of the Milky Way models. The fastest, median, and slowest Sgr orbits correspond to the first, fifth, and ninth bin (out of nine bins). One other consideration in this section is that the slowest Sgr orbit in MWP14-1 is only integrated back one apocentre before being placed. This is because the second last pericentre occurred $\sim8$ Gyr ago and all effects from that apocentre would have phase mixed, so in that case only, the shape of the perturbation is set only by the most recent pericentre passage. In the MWP14-1 simulation the number of particles in each solar neighbourhood of the different simulations from Fastest to Slowest is: 489\,542, 571\,458, and 803\,395. For MWP14-3, the solar neighbourhoods contain 962\,493, 1\,026\,534, and 949\,064 particles from Fastest to Slowest.

\begin{figure}
    \centering
    \includegraphics[width=0.5\textwidth]{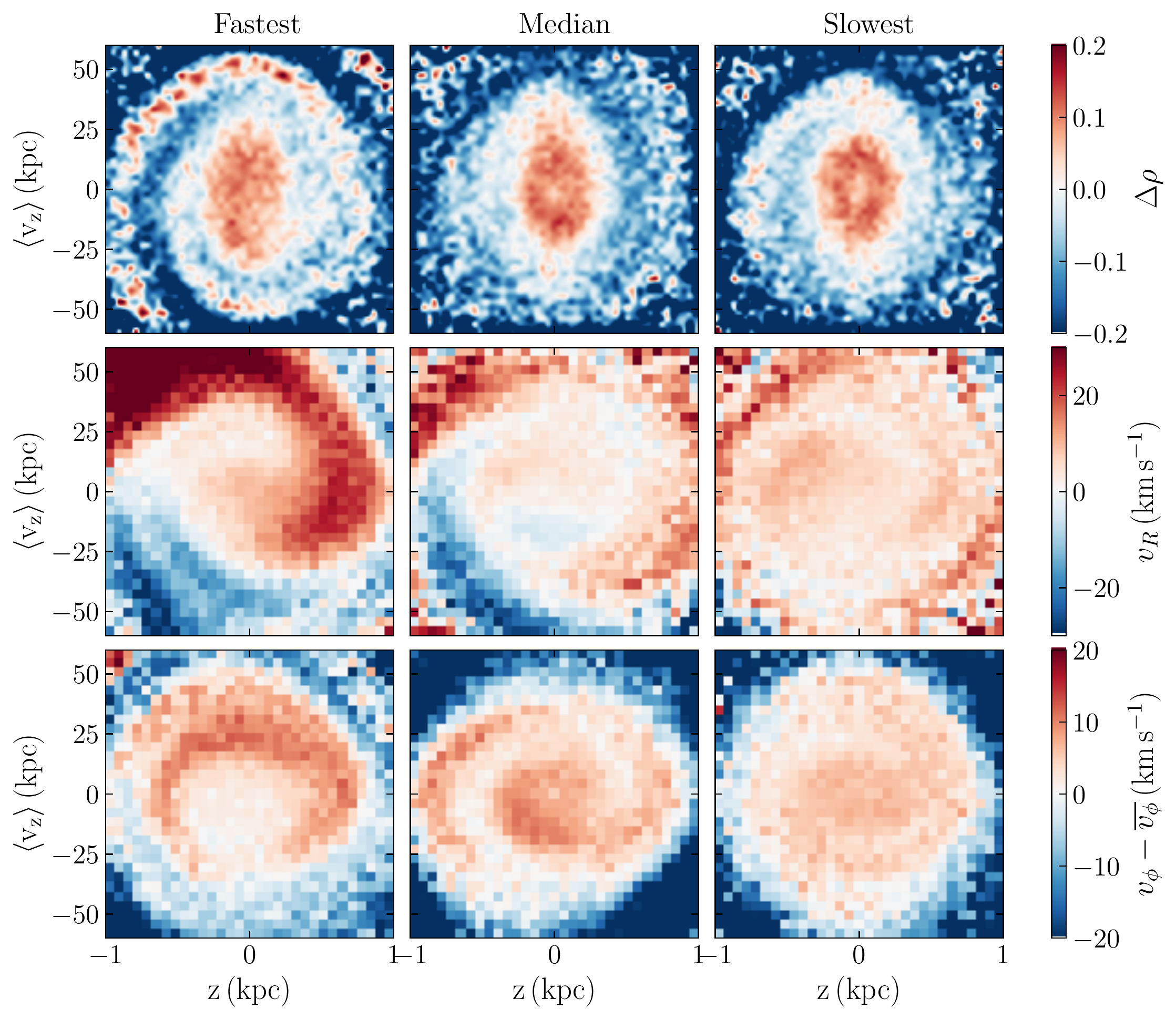}
    \caption{The same as \figurename \ref{fig:phase_MW1}, but for MWP14-3, the initial condition with the heaviest Milky Way halo. }
    \label{fig:phase_MW3}
\end{figure}

In BB21 we found that the mass of the Milky Way had a significant effect on the shape and amplitude of the perturbation, especially the asymmetry, which we wanted to follow up with $N$-body simulations. Investigating two more Milky Way models requires new initial conditions to be set up and run to equilibrium. Due to computational constraints, we decided to halve the number of particles in our MWP14-1 and MWP14-3 simulations; therefore all simulations in this section have approximately 500 million particles. The initial conditions for both are initialised using \texttt{GalIC} as described in \secname \ref{sec:MWmodel} using the values from \tablename \ref{tbl:MWGalIC}. We chose to run simulations with all three Milky Way mass halos for completeness to ensure we covered a reasonable range of possible Milky Way halo masses, though we expect MWP14-2 or MWP14-3 to provide the best match based on the results of BB21. For MWP14-3, we found that the outer reaches of the disc took longer to reach equilibrium in our isolated Milky Way simulation, so the equilibrium simulation was run for 5 Gyr instead of 3 Gyr. 

In Figure \ref{fig:MWmass_asym_meanV}, we show the vertical density profile, asymmetry and mean vertical velocity for the three different Milky Way models for each of the three different Sgr velocity bins.  Looking first at the density profiles, we see that the mid-plane density of MWP14-1 is much lower than the other two simulations in both the Fastest and Median simulations. This likely happens because the disc in the MWP14-1 experiences a smaller restoring force due to the smaller halo mass and is not able to compete with the large force from the Fastest and Median Sgr encounters, and is therefore more fluffed up by the encounter. As previously discussed, the varying mid-plane densities will greatly affect any possible trends we could see, and therefore we will only discuss MWP14-1 in terms of trends for the Slowest simulation. In the Fastest simulation, the amplitude of the perturbation in MWP14-2 is quite large and does not follow the same oscillatory pattern as seen in other smaller scale perturbations. This is likely due to the exact configuration in MWP14-2 leading to a closer passage by the solar neighbourhood than the other two simulations. In BB21, we found that as we increased the Milky Way halo mass in general, Sgr passed through the disc at higher velocity and more recently and the asymmetry wavelength and the asymmetry amplitude increased. This was not consistent across all Sgr models and all Sgr kinematics, but held in most cases. The outliers were due to closer encounters between Sgr and the solar neighbourhood due to a combination of the circular velocity of the neighbourhood and the effects of dynamical friction in that specific halo. It is therefore also difficult to use the Fastest MWP14-2 simulation to uncover trends. In the Median simulation, all three perturbations have a much more realistic amplitude. Again, we exclude MWP14-1 due to the small mid-plane density. However, we do see that the amplitude of the asymmetry in MWP14-3 is generally larger than the amplitude of the perturbation in MWP14-2. Furthermore, we see that within 2 kpc, the asymmetry in MWP14-3 peaks twice while the asymmetry in MWP14-2 peaks three or more times. This tells us that as we increase the Milky Way halo mass, the amplitude and the asymmetry wavelength both increase, in agreement with our expectations from BB21. Finally, in the slowest Sgr orbit, with all of the mid-plane densities being fairly similar, we are able to compare across all three Milky Way halo masses. Clearly, the amplitude of the asymmetry increases as we increase the mass of the halo as does the perturbation wavelength, though the exact asymmetry wavelength in MWP14-1 is difficult to discern given the very small amplitude of the perturbation. 

We next consider the mean vertical velocity for each of the simulations. Across all nine simulations considered, the Fastest MWP14-1 simulation is the only one with a clear bending mode. The locations of the troughs are similar to that those seen in the data, however the amplitude is much too small. Furthermore, the asymmetry for that simulation was far from a match to the data. With less particles in the solar neighbourhood-like volumes, the mean vertical velocity measurements for MWP14-1 and MWP14-3 come with much larger uncertainties. So, while it is difficult to uncover trends in the mean vertical velocity, it is easy to tell that none are within uncertainty of the true observations, especially the dip at $z\sim0.5$ kpc. While recovering oscillatory asymmetry measurements in our simulation is quite easy, it seems that recovering mean vertical velocity perturbations on the order of a couple km\,s$^{-1}$ is much more difficult using Sgr. 

In BB21, we found that MWP14-3 resulted in the perturbations with the most accurate shape, but consistently found that the asymmetry wavelength was too large. In our simulations, the model with the asymmetry and mean vertical velocity most similar to the \Gaia DR2 observations is the median velocity Sgr in MWP14-3, the heaviest halo. The asymmetry for that simulation is the best match for the dip in the asymmetry at $z\approx0.4$ kpc, but it still far from an exact match. The dip in the asymmetry of the simulation is centred closer to $z=0.7$ kpc. Furthermore, the peak in the simulation asymmetry at $z\approx1.0$ is not seen in the \Gaia DR2 data. If we calculate the breathing and bending mode amplitudes for that simulation, we find that the bending and breathing amplitudes of are within uncertainty of each other at all heights. This tells us that the symmetric vs. antisymmetric components of the velocity are approximately equal. In \figurename C.6 of \cite{GaiaKinematics} we see that there should not be any amplitude of a breathing mode at the Sun's location out to 1.2 kpc, let alone a breathing mode with approximately the same amplitude as the bending mode. Furthermore, the final mass of Sgr in that simulation is approximately $15\times 10^8$ M$_\odot$, which is approximately 3-6 times too large compared to the measured mass of the Sgr remnant \citep{Vasiliev2020} and yet the amplitude of the dip in the asymmetry near $z=0.5$ kpc is still too small. Therefore, while we can conclude that the median Sgr velocity model for MWP14-3 is the best at reproducing one feature of the asymmetry, it is not a good match.

Finally, we also look at the phase space for each of the simulations coloured by the normalized change in density, the radial velocity, and the normalized azimuthal velocity. \figurename \ref{fig:phase_MW1} shows the phase-space spirals seen in the MWP14-1 simulations. There are no discernible trends in the phase space when comparing the different Sgr orbit speeds. In fact, for the Slowest simulation, it is difficult to resolve a spiral at all in the normalized density due to the low amplitude and low number of particles. In the other two simulations, we can compare the phase-space spiral to the one seen in the \Gaia data, and again find that they are too loosely wound to be a match. One particularly interesting feature is the very distinct spiral in the Fastest simulation when coloured by the radial velocity. This is surprising because that choice of colouring phase space is the least distinct in the data, yet is very clear in our simulation. \figurename \ref{fig:phase_MW3} shows the phase-space spiral for the three simulations run in MWP14-3. Again, it is difficult to discern any patterns between the simulations. However, one thing of note is the Median simulation when coloured by the normalized azimuthal velocity. In the \Gaia data, there is clearly only one arm that spirals around. In our simulation, it appears there are two distinct arms spiralling out from the centre. This is further proof that while the asymmetry of the Median simulation in MWP14-3 might have vaguely resembled observations, it is far from a match, leaving us once again with no consistent match to observations throughout all of our simulations. 

\section{Discussion \& Conclusion}\label{sec:conc}

The origin of the phase-space spiral has been a largely debated topic. One of the leading theories is that the passage of Sgr caused oscillations in the solar neighbourhood. This paper focuses on using a suite of 13 simulations that explore the parameter space of Sgr's mass and orbit, as well as the Milky Way halo mass, to help address this question. These simulations are intended to be shared with the research community for further investigations. 

We start by describing a thorough method for initializing Milky Way initial conditions derived from the properties of \texttt{MWPotential2014} in \texttt{galpy}. We consider three different Milky Way initial conditions with varying halo masses that range from $0.8-1.6\time10^8$ M$_\odot$. Once we had the Milky Way initial conditions, we evolved each galaxy for 3+ Gyr in isolation using an $N$-body GPU tree code \texttt{Bonsai}. We did this to ensure that the solar neighbourhood was in fact in equilibrium and that the dominant effects we saw would be from the passage of Sgr. In all three galaxies, a bar formed in the centre of the galaxy, much like it did in the Milky Way. We then looked at the properties of the final snapshot of the equilibrium galaxies, including the asymmetry at eight different solar neighbourhood-like volumes around the simulated Milky Ways and found that the intrinsic asymmetry of the disc is smaller than $\sim0.025$ out to 1.5 kpc, and smaller than $\sim0.1$ between $1.5$ kpc $<z<2$ kpc.

Once we had the initial condition, we had to figure out where to place the Sgr particles such that they would finish near the true Sgr position. In our simulations, we use three different two-component Sgr models in our simulations with total initial masses ranging from $(5.1-51)\times10^9$ M$_\odot$. Placing the Sgr particles is fairly complicated as it involves estimating the Milky Way potential, mass loss of Sgr, and dynamical friction. We solved the first dilemma by calculating an SCF potential expansion of the $N$-body particles to estimate the true potential of the unperturbed simulation. To take into account the changing mass of Sgr in the calculation of dynamical friction, we estimated the mass loss of Sgr by running a 100 million particle simulation of Sgr and tracking the virial mass throughout the simulation. Finally, we used a Chandrasekhar dynamical friction object in \texttt{galpy} to estimate the dynamical friction of the orbit. This allowed us to estimate where to place Sgr in our simulations. Table \ref{tbl:SgrPos} shows the final position of Sgr in all of our simulations as well as the 'True' position.

In each of our simulations, we investigated three different properties of the simulation to see how it affected the perturbation to the solar neighbourhood. The first was the mass of Sgr. We found that as we decreased the mass of Sgr, the both the amplitude and the wavelength of the asymmetry also decreased. Looking at the mean vertical velocity, we found that all of the models produced a breathing mode, unlike what is seen in the \Gaia data. Finally, we looked at the phase space for each of the three simulations and found that the spiral in the normalized density change is much less tightly wound than what has been observed in the solar neighbourhood. By colouring phase space by the radial velocity, the amplitude of the radial velocity across phase space decreases as the mass of Sgr decreases. Finally, when coloured by the normalized azimuthal velocity, we also find that we can see the spiral become more tightly wound as the Sgr mass decreases. Between the shape of the asymmetry, the breathing mode in the mean vertical velocity, and the loosely wound spiral we consistently find that we are unable to match the observations. We can therefore conclude that considering changes to Sgr's mass is not enough to match observations.

Next, we looked at the effect of the uncertainty in Sgr's current-day position on the observed perturbation. With such a large uncertainty in Sgr's current position, there is also a large uncertainty in its orbit. We characterise an orbit by the speed at which it passed through the disc most recently, $v_{z,\mathrm{through}}$. We chose to compare four different orbits: Fastest, Fast, Median, Slow, and Slowest. For these simulations, we chose to use the medium mass model of Sgr in the middle halo mass, MWP14-2. Comparing the simulations, we found in both the asymmetry and the phase-space spiral that the wavelength of the perturbation and the perturbation amplitude decrease as the velocity of Sgr decreases. We found that none of the simulations were able to reproduce an asymmetry, mean vertical velocity, or phase-space spiral similar to that observed in \Gaia DR2. The most similar in shape was the Fast simulation, but its amplitude was too small and the asymmetry wavelength was too long. Furthermore, it produced a breathing mode in the mean vertical velocity, which is not seen in the data. Considering the trends seen in our simulations, it is unlikely that changing the speed of Sgr is enough to remedy the discrepancy between the Sgr's effect on the neighbourhood and observations.

Finally, we looked at how the Milky Way halo mass affects the observed asymmetry. While the sections on Sgr mass and Sgr kinematics used the medium mass halo for the simulations, we also looked at a light and a heavy halo. Comparing the different mass halos, we discovered that as we increase the Milky Way halo mass, the amplitude and the asymmetry wavelength both increase. In the light halo simulation, we found that none of the perturbations were similar to the observations from \Gaia DR2. Furthermore, the small \midplane density in the simulation also meant that the both the amplitude and wavelength of the asymmetry were overestimated, making the simulations an even worse match to observations. The heavy Milky Way halo simulation resulted in the best match of all the simulations considered. The simulation where Sgr was initialized with the median $v_{z,\mathrm{through}}$ resulted in a dip in the asymmetry at approximately $z=0.7$ kpc. However, the asymmetry for that simulation also contained a peak at $z=1.0$ kpc that is not seen in the data. Furthermore, the mean vertical velocity contains a measurable breathing mode within $z<1$ kpc that does not exist in the \Gaia DR2 data. This is also seen in the phase-space spiral that has two distinct arms (breathing) instead of one (bending) \citep{Hunt2021}. We therefore conclude that none of the simulations are able to reproduce the observed perturbation to the solar neighbourhood. Furthermore, we found across the board that the Sgr progenitor remnant was consistently much heavier $(10-40\times10^8$ M$_\odot)$ than the true estimate values of Sgr $(2.5-5\times10^8$ M$_\odot)$.

Our suite of simulations covers a larger range of parameter space in the Sgr-Milky Way interaction than has previously been explored in one study. Thirteen simulations of the Milky Way and Sgr interaction is fairly numerous, but while we tried to cover the range of each parameter, our sampling within that range was fairly small. A further limitation of our analysis is that we are not able to control the \midplane density of the solar neighbourhood. This means we cannot force it to exactly match the mid-plane density of the Milky Way disc, something BB21 shows can affect the perturbation wavelength and amplitude. The lack of a gas disc is also a limitation of our simulations, though it is a fairly common simplification \citep{Laporte2018,Hunt2021} and simply assumes that the gas approximately follows the stars. Furthermore, hydrodynamical simulations are currently unable to match the resolution achieved in these simulations, which means we would be unable to resolve the solar neighbourhood features to the accuracy required. Finally, from our analysis in \secname \ref{sec:MWmass}, we see that the mass of the halo, and therefore dynamical friction, has a large effect on the observed perturbation. Our simulations span halo masses of $(0.98-1.95)\times10^{12}$ M$_\odot$, which is a very large range to be covered by only three simulations. In our analysis, we were only able to look at three different Milky Way mass halos especially since they each had to be integrated for 3 or more Gyr each. Finally, we do not consider cuspy halos. However, they are more resilient to stripping and would therefore maintain a larger mass by the end of our simulation. Our Sgr models were already too large, so it is unlikely that a cuspy halo would lead to a more accurate final mass of Sgr. Alternatively, as previously discussed, some studies have suggested that Sgr must lose mass more quickly than what is seen in our simulations \citep{BH2021}. While this may improve the disagreement in the amplitude of the simulations and observations, it is unlikely to resolve the discrepancy in the shape and wavelength of the perturbation. While the span of the simulations is larger than has been considered before, there are still several areas of improvement. 

Despite these limitations, we can confidently rule out Sgr as the dominant cause of the perturbations to the solar neighbourhood. We suggest that with a much smaller mass than utilised here, it is likely that any perturbation from Sgr would be a secondary effect on the solar neighbourhood and certainly not the dominant one as has previously been suggested. First, the final mass of Sgr in all of our simulations is significantly larger than the Sgr mass estimated by \cite{Vasiliev2020} as shown in \tablename \ref{tbl:SgrPos}. Despite them all having too large a mass, we still consistently find that the amplitudes within $|z|<1$ kpc are too small with the exception of a couple of the Fastest models or the Heaviest Sgr, but none of which have the correct shape or mean vertical velocity or phase-space spirals. In fact, we find that all of our simulations have a spiral that is much too loosely wound, a common dilemma with Sgr simulations \citep{Gomez2013,Laporte2019}. Our mass models of the Milky Way halo spans halo masses of $(0.8-1.6)\times10^{12}$ M$_\odot$ that largely covers the range of possible masses in literature \citep{Callingham2019}, so it is unlikely that the solution to the discrepancy is a heavier or lighter halo than considered here. Even our best match from the simulations only had one feature that vaguely resembled the observed perturbations, but overall was a poor match. Much like our investigation in BB21, we conclude that Sgr could not have caused the perturbation of the solar neighbourhood. We cannot rule out the possibility that Sgr might play some role in the perturbation, such as a spiral+bar+Sgr coupling, but it likely plays a secondary role. The consistent discrepancy in the perturbation wavelength and the shape of the perturbation across all simulations, especially in the phase space, leads us to conclude that Sgr cannot be the driving mechanic for the perturbation to the solar neighbourhood. It is highly unlikely that any `fine-tuning' of the Sgr parameters will result in a match to observations given our large range of values as well as the extensive number of simulations performed by other authors \citep[e.g.,][]{Laporte2018,BH2021}. It remains to be seen whether other proposed scenarios such as the buckling of the bar, or nonlinear coupling between bar-buckling, spiral structure, and perturbations from satellites including Sgr can explain the vertical asymmetry in the solar neighbourhood.

\acknowledgments

We thank the anonymous referee for a helpful referee report. MB and JB acknowledge financial support from NSERC (funding reference number RGPIN-2020-04712) and an Ontario Early Researcher Award (ER16-12-061). JASH is supported by a Flatiron Research Fellowship at the Flatiron institute, which is supported by the Simons Foundation. This work has made use of data from the European Space Agency (ESA) mission
{\it Gaia} (\url{https://www.cosmos.esa.int/gaia}), processed by the {\it Gaia}
Data Processing and Analysis Consortium (DPAC,
\url{https://www.cosmos.esa.int/web/gaia/dpac/consortium}). Funding for the DPAC
has been provided by national institutions, in particular the institutions
participating in the {\it Gaia} Multilateral Agreement. A portion of the computations were performed on the Niagara supercomputer at the
SciNet HPC Consortium \citep{Loken2010,Ponce2019}. SciNet is funded by: the Canada Foundation for Innovation; the Government of Ontario; Ontario Research Fund -
Research Excellence; and the University of Toronto. This work was performed in part at Aspen Center for Physics, which is supported by National Science Foundation grant PHY-1607611. This work was partially supported by a grant from the Simons Foundation.

\bibliography{references}{}
\bibliographystyle{aasjournal}

\end{document}